\documentclass[11pt,nofootinbib,superscriptaddress]{revtex4-2}
\usepackage[nohead]{geometry}
\usepackage{physics}
\usepackage{hyperref}
\usepackage{graphicx} % Allows for eps images
\usepackage[svgnames]{xcolor} % Required to specify font color
\usepackage{amsmath} % AMS Math Package
\usepackage{amssymb}	% Math symbols such as \mathbb
\usepackage{slashed}
\usepackage{tikz-feynman}
\usetikzlibrary{decorations.pathmorphing}
\usepackage{subcaption}
\usepackage{float}
\begin{document}

%\date{\today}

\title{Massive photon propagator in the presence of axionic fluctuations}

\author{B. A. S. D. Chrispim}
\email[E-mail:]{chrispim.breno@posgraduacao.uerj.br}
\affiliation{Departamento de F\'{\i }sica Te\'{o}rica, Instituto de F\'{\i }sica, UERJ - Universidade do Estado do Rio de Janeiro.\\ Rua S\~{a}o Francisco Xavier 524, 20550-013 Maracan\~{a}, Rio de Janeiro, Brasil.}
\author{R. C. L. Bruni}
\email[E-mail:]{bruni.r.c.l@gmail.com}
\affiliation{Departamento de F\'{\i }sica Te\'{o}rica, Instituto de F\'{\i }sica, UERJ - Universidade do Estado do Rio de Janeiro.\\ Rua S\~{a}o Francisco Xavier 524, 20550-013 Maracan\~{a}, Rio de Janeiro, Brasil.}
\author{M. S. Guimaraes}
\email[E-mail:]{msguimaraes@uerj.br}
\affiliation{Departamento de F\'{\i }sica Te\'{o}rica, Instituto de F\'{\i }sica, UERJ - Universidade do Estado do Rio de Janeiro.\\ Rua S\~{a}o Francisco Xavier 524, 20550-013 Maracan\~{a}, Rio de Janeiro, Brasil.}

%----------------------------------------------------------------------------------------
%	ABSTRACT
%----------------------------------------------------------------------------------------
\begin{abstract}
The theory of massive photons in the presence of axions is studied as the effective theory describing the electromagnetic response of semimetals when a particular quartic fermionic pairing perturbation triggers the formation of charged chiral condensates, giving rise to an axionic superconductor. We investigate corrections to the Yukawa-like potential mediated by massive photons due to axion excitations up to one-loop order and compute the modifications of the London penetration length.
\end{abstract}

\maketitle

%----------------------------------------------------------------------------------------
%	Introduction
%----------------------------------------------------------------------------------------
\section{Introduction}

The origin of axion physics can be traced to the existence of the quark chiral condensate in QCD. Chiral spontaneous symmetry breaking leads to the naive prediction of certain quasi-Goldstones bosons associated with the $U(1)$ chiral symmetry that does not materialize in observations \cite{Weinberg:1975ui}. 't Hooft \cite{tHooft:1976snw, tHooft:1980xss, tHooft:1986ooh} was able to explain away these spurious particles observing that the chiral anomaly could lead to an explicit symmetry breaking (as opposed to spontaneous) due to instantons contributions, thus solving the $U(1)$ problem. But, once instantons are considered, one has to deal with the ensuing violation of parity $P$ and time-reversal $T$ symmetries associated with the $\theta$ term $\sim \theta \tilde{F} F$. The lack of observational proof of these symmetry violations in QCD experiments is historically known as the strong CP problem since charge conjugation $C$ is preserved. In order to make sense of this, one has to fine-tune the offending $\theta$ parameter to be sufficiently small. A solution to this undesirable fine-tuning was proposed by Peccei and Quinn \cite{Peccei:1977hh, Peccei:1977ur} (see \cite{Peccei:2006as} for a review) that promoted the parameter $\theta$ to a dynamical field introducing an associated abelian global symmetry, dubbed $U(1)_{PQ}$ by Weinberg \cite{Weinberg:1977ma}, and a new particle, a pseudoscalar named Axion by Wilczek \cite{Wilczek:1977pj}. Since then there have been many investigations, both theoretically and experimentally \cite{Kim:2008hd, Braaten:2019knj}, of this hypothetical particle. Even though the original Axion construction of Peccei-Quinn-Weinberg-Wilczek is ruled out by experiments there have been other constructions demanding different extra fields such as the ``invisible Axion'' models that are still alive as viable options \cite{Kim:1979if, Shifman:1979if, Dine:1981rt}. Axion physics has been revisited time and again over the years upon the expectation that it can serve as a good description of a variety of phenomena. Most notably it has been associated with a promising candidate for dark matter components having relevant contributions to cosmology (see \cite{Marsh:2015xka} for a review).

Theories constructed with Axion-like particles have some ``universal'' properties due to their unique coupling with the gauge fields. For example, $P$ and $T$ symmetry breaking and the sensibility to the topological structure of gauge fields, exemplified by instantons in the QCD context, makes it clear that this kind of coupling is bound to show up in effective field theories that share these properties. Another aspect of this is the fact that Axion-like particle will couple to any gauge field with respect to which the anomalous fermions have charge, since this is a consequence of the chiral transformation of the integration measure (Fujikawa method). In QCD, for instance, Axions couple with the gluon fields and also with the electromagnetic field, since quarks are electrically charged. This gave rise to the study of Axion electrodynamics phenomenology \cite{Wilczek:1987mv} leading to some interesting insights about deformations in the electromagnetic wave propagation as a source for detection of astrophysics signature of Axions. Recently, a whole new avenue for investigations was opened steaming from the discovery of topological materials \cite{2010Natur.464..194M, Hasan:2010xy, Qi:2011zya, Hasan2011ThreeDimensionalTI}. Most of these materials display a nontrivial response under $P$ and $T$ transformation. Also, effective emergent chiral symmetries appear in their mathematical modeling, which has been shown to lead to the unavoidable introduction of effective axion-like excitations. The curious behavior of axion electrodynamics \cite{Wilczek:1987mv} has encountered numerous applications in condensed matter phenomenology of topological materials, playing an important role in the effective description of the electromagnetic response in those systems, where axionic couplings have appeared in many guises. 

The preceding discussion led us to believe that is necessary to investigate further the interplay between Axion-like excitations and gauge field dynamics. To this end, we will focus on the phenomenology of topological superconductors by constructing an effective theory in a Dirac semimetal with quartic interaction. The result is an abelian Proca field theory with axion-like interaction. We will study the resulting modifications in the propagation of the massive vector particle when subject to axion-like fluctuations by computing the $1$-loop corrections to the two-point function of the Proca field.

This work is organized as follows: In section II we motivate the model by relating it to an effective description of a superconductor obtained by perturbing a Dirac semimetal with a four fermion interaction. In section III we define our notation and the action of the model with all its coefficients and renormalization factors. This will set the stage for the discussion of the (massive) photon self-energy in section IV. In section V we present our main results concerning the modified Yukawa potential between static charges induced by the axion dynamics. The analysis of the results are discussed in section VI and the limit of large relative masses, and the connection with the phenomenology of London's length, is examined as well. Finally, in section VII we present our conclusions and the appendix provides some details of the computation.

%----------------------------------------------------------------------------------------
%	A superconducting model from Dirac semimetals
%----------------------------------------------------------------------------------------
\section{A superconducting model from semimetals}

The introduction of axion-like interaction for the effective electromagnetic description of topological materials was developed in \cite{Qi:2008ew} for the case of topological insulators. The non-trivial phenomenology originates from a spacetime dependent Axion-like field, as can be seen from the modified Maxwell's equations
\begin{subequations}\label{Maxwell-Axion}
\begin{align}
\div {\bf E} &= \rho - \frac{e^2}{4\pi^2} \grad\theta\cdot {\bf B}\\
\grad \times {\bf B} &= {\bf j}+\frac{\partial {\bf E}}{\partial t} +  \frac{e^2}{4\pi^2} \left(\grad\theta \times {\bf E} + \frac{\partial \theta}{\partial t} {\bf B} \right)\\
\div {\bf B} &= 0\\
\grad \times {\bf E} &= - \frac{\partial {\bf B}}{\partial t}
\end{align}
\end{subequations}
Here, a normal insulator is characterized by $\theta = 0$ $(\text{mod }2\pi)$ while a topological time-reversal invariant insulator is described by having $\theta = \pi$ $(\text{mod }2\pi)$. The interface between these two phases must be a smooth transition between the two defining values of $\theta$, so one expects a spacetime varying Axion field interpolating between $0$ and $\pi$ where the dynamics are described by Axion-Maxwell electromagnetism \eqref{Maxwell-Axion}. This setting describes various phenomena, v.g. a constant magnetic field leads to a charge density proportional to the applied field. Also, there is the possibility of currents with components perpendicular to an applied external electric field (Quantum Hall effect \cite{quantum.hall.effect.40.years}) and parallel to an external magnetic one (chiral magnetic effect \cite{Fukushima_2008}), both with a quantized proportionality coefficient.

Axion-like terms are also relevant for the description of Weyl semimetals \cite{2018RvMP...90a5001A, 2017ARCMP...8..337Y}, i.e. systems whose band structure intercepts at two or more points in momenta space around which a linear dispersion approximation is valid. This description leads to fermionic excitations with a definite helicity, that is, projection of the spin along the momentum direction, thus defining Weyl fermions. Helicity coincides with chirality for massless fermions and the chirality of these excitations is measurable by the flux of the Berry curvature in the Brillouin zone. Furthermore, for topological reasons, the total flux must be zero inside a Brillouin's zone (Nielsen-Ninomyia theorem \cite{Nielsen:1981hk, Friedan:1982nk}), which explains why Weyl fermions always appear in pairs of opposite chirality. When two Weyl fermions are at the same point in momentum space, they build up a Dirac fermion, which arises, for example, in the description of the electronic structure of graphene (a type of Dirac semimetal). 
Experimental investigations of Weyl metals have been undertaken. It was shown, for instance, in \cite{PhysRevLett.111.246603} that  Weyl fermions appear in $Bi_{x-1}Sb_x$ near the critical point of the topological phase transition when magnetic fields are applied.

An interesting setting occurs when two Weyl points are separated in momentum and energy but are close to the Fermi surface. The theoretical description of this situation can be conveniently expressed by a Dirac action where the right and left Weyl modes are arranged on a Dirac spinor $\psi = \smqty(\psi_L\\\psi_R)$  with $\bar{\psi} = \psi^{\dagger}\gamma^0 = \smqty(\psi_R^{\dagger}&\psi_L^{\dagger})$
\begin{align}
\label{dirac}
S = \int \dd[4]{x} \bar{\psi}(x) \left(i\slashed{\partial} + \slashed{b}\gamma^5 + ie\slashed{A}(x) \right)\psi(x),
\end{align}
and an interaction with an external electromagnetic gauge potential $A_{\mu}$ was also included. The $4$-vector $b_{\mu}$ is constant and represents the separation in the energy-momentum space of the Weyl points. Chirality of the Weyl components means that $\gamma^5\smqty(\psi_L\\\psi_R)=\smqty(-\psi_L\\\psi_R)$ and thus one can clearly note in \eqref{dirac} that left-handed and right-handed fermions are shifted in opposing directions along $b_{\mu}$ in energy-momentum space. As described in details in \cite{2012PhRvB..86k5133Z}, one can eliminate $b_{\mu}$ by performing a (local) chiral transformation
\begin{align}
\label{chiraltranf}
\psi(x) \rightarrow e^{i\frac{1}{2}\theta_0(x)\gamma^5} \psi(x)
\end{align}
with $\theta_0(x) = 2b_{\mu} x^{\mu}$. This is, of course, not a symmetry of the action, but just a change in the fermionic variables. In the quantum path integral formulation, this transformation gives rise to a non-trivial contribution from the jacobian of the fermionic integration measure, well known from the chiral anomaly. Thus the effective action for the electromagnetic response becomes 
\begin{align}
\label{diracAxion}
S \rightarrow \frac{e^2}{32\pi^2} \int \dd[4]{x}\theta_0(x) \varepsilon^{\mu\nu\rho\sigma}F_{\mu\nu}F_{\rho\sigma} - i\ln \det \left(i\slashed{\partial} +  ie\slashed{A}(x) \right),
\end{align}
where $F_{\mu\nu} = \partial_{\mu}A_{\nu} - \partial_{\nu}A_{\mu}$. So, in essence, the Weyl semimetal system naturally displays an Axion-like term that encodes the energy-momentum separation of the Weyl nodes. This term is responsible for the phenomenology described by equations \eqref{Maxwell-Axion}. In this particular setting, the axion-like field has linear spacetime dependency that leads to a constant \textbf{external} $4$-vector that was thoroughly studied in the context Lorentz violating field theories \cite{Carroll:1989vb}.

One can go further and consider the case where the Axion-like field is dynamical. As pointed out in \cite{Wang:2012bgb, 2014PhRvB..90c5126M, You:2016wbd}, this seems to be a fruitful endeavor since chiral symmetry can be dynamically broken due to the formation of a chiral condensation induced by the four fermions pairing interaction
\begin{align}
\label{pairing}
\lambda^2 \left( \bar{\psi}(x) P_L \psi(x)\right) \left( \bar{\psi}(x) P_R \psi(x)\right)
\end{align}
where $P_L = \frac{1}{2}(1-\gamma^5)$ and $P_R = \frac{1}{2}(1+\gamma^5)$ are chiral projectors and the coupling $\lambda$ has mass dimension $-1$. Note that this pairing connects left and right handed fields. In fact, $\bar{\psi}(x) P_L \psi(x) = \psi^{\dagger}(x) P_R \gamma^0 P_L\psi(x) = \psi_R^{\dagger}(x)\psi_L(x)$. \footnote{Throughout this paper we use the van der Waerden notation of dotted and undotted spinor indexes: $\psi_{L\alpha}$ and $\psi_{R}^{\dot{\alpha}}$  are the left and right spinors. Spinor index contractions are defined as $ \psi_R^{\dagger}(x)\psi_L(x) = \psi_R^{\dagger \alpha}(x)\psi_{L\alpha}(x) =\psi^{\dagger}_{R\alpha}(x)\varepsilon^{\alpha\beta}\psi_{L\beta}(x) $. And similarly for other billinears we shall encounter, for instance,  $\psi_R(x)\psi_R(x) = \psi_{R \dot{\alpha}}\psi_{R}^{\dot{\alpha}} = \psi_{R}^{\dot{\alpha}}\varepsilon_{\dot{\alpha}\dot{\beta}}\psi_{R}^{\dot{\beta}} $ and $\psi_L(x)\psi_L(x) = \psi_{L}^{\alpha}(x)\psi_{L\alpha}(x) =\psi_{L\alpha}(x)\varepsilon^{\alpha\beta}\psi_{L\beta}(x) $.}
One can formulate the description of the system by including this four fermion interaction in \eqref{dirac}, written with the help of a Hubbard-Stratanovich auxiliary complex field $\Phi(x)$         
\begin{align}
\label{diracMass}
S &= \int \dd[4]{x} \bar{\psi}(x) \left[ \left(i\slashed{\partial} + \slashed{b}\gamma^5 + ie\slashed{A}(x) - \lambda\Phi(x) \frac{1}{2} \sigma^0\otimes \left(\tau_1 + i \tau_2\right)  + \lambda\Phi^{\dagger}(x) \frac{1}{2}\sigma^0\otimes \left(\tau_1 -  i \tau_2\right)\right)\psi(x) + |\Phi(x)|^2 \right],
\end{align}
where we introduced the matrix structure $\sigma \otimes \tau$, such that $\sigma$ and $\tau$ are Pauli matrices  ($\sigma^0$ is the identity) acting on spin degrees of freedom and helicity, respectively. The auxiliary field is determined by its extrema in the action and results in
\begin{align}
\label{condensate}
 \Phi(x) = \lambda  \bar{\psi}(x) P_L \psi(x) = \lambda \psi_R^{\dagger}(x)\psi_L(x)
\end{align}
It is argued in \cite{Wang:2012bgb, 2014PhRvB..90c5126M} that the strong coupling dynamics of the theory favors the formation of a condensate $\langle \Phi \rangle \neq 0$, resulting in the dynamical break of the chiral symmetry following the Peccei-Quinn mechanism. In this context, small fluctuations around the condensate $\langle \psi_R^{\dagger}(x)\psi_L(x) \rangle = v^3$ can be approximated by
\begin{align}
	\label{condensate2}
	\Phi(x) = \lambda v^3e^{i\frac{\theta(x)}{f}}  
\end{align}
where $f$ is a mass scale and $v^3$ has mass dimension $3$. After redefining the fermion field $\psi(x) \rightarrow e^{-i\frac{1}{2}\qty(\theta_0(x) + \frac{\theta(x)}{f})\gamma^5} \psi(x)$, where (again) $\theta_0(x)$ was included to cancel the $b$ term. Finally, taking into account the Jacobian of the transformation, the effective action becomes
\begin{align}
\label{diracAxionfluct}
S \rightarrow \frac{e^2}{32\pi^2} \int \dd[4]{x}\qty(\theta_0(x)+\frac{\theta(x)}{f}) \varepsilon^{\mu\nu\rho\sigma}F_{\mu\nu}F_{\rho\sigma} - i\ln \det \left(i\slashed{\partial} 
+i\gamma^5\frac{\slashed{\partial}\theta(x)}{f}+  ie\slashed{A}(x) + \lambda^2 v^3 \right)
\end{align}
This effective electromagnetic theory displays a dynamical axion-like field $\theta(x)$, its bilinear kinetic term originates from the derivative expansion of the fermionic determinant and set to the canonical form by imposing $f\sim \lambda^2 v^3$. Furthermore, the condensate provides a mass for the axion of order $\frac{\lambda v^3}{f} \sim \frac{1}{\lambda}$, which is analogous to the charge density waves.
\begin{align}
\label{CDW}
\langle \bar{\psi}\psi \rangle =  \langle\bar{\psi}(x) P_L \psi(x) \rangle + \langle\bar{\psi}(x) P_R \psi(x) \rangle = \frac{1}{\lambda}\left(\langle\Phi(x) \rangle + \langle\Phi^{\ast}(x) \rangle\right) \sim 2v^3 \cos\left(\theta_0(x) +  \frac{\theta(x)}{f}\right)
\end{align}
The resulting effective theory is the same proposed as a description of a topological magnetic insulator in  \cite{2010NatPh...6..284L}. This signals a possible transition from Weyl semimetals to topological magnetic insulators induced by the vacuum instability resulting from the four fermions interaction. 

The pairing just discussed establishes an inter-node connection that breaks chiral symmetry resulting in an electromagnetic theory with axionic fluctuations. Following this idea, in order to construct a superconducting state with axionic fluctuations, it is necessary to seek a pairing that breaks charge symmetry and chiral symmetry. The important question about the leading mechanism for the superconducting instability and the different pairings that can lead to it in a Weyl semimetal system has been a subject of intense investigation during the last few years. Pairings such as intra-node FFLO pairing \cite{PhysRevB.92.035153, PhysRevLett.124.096603, PhysRevB.86.214514}, which involves a nontrivial center-of-mass momenta dependence \cite{PhysRev.135.A550} and inter-node BCS pairing \cite{PhysRevB.86.214514, PhysRevB.89.014506, PhysRevLett.120.067003}, which connects fermionic excitations in the opposite Fermi-surfaces, and therefore with opposite chiralities, have attracted attention. More general BCS-like pairings, like the triplet \cite{PhysRevB.89.014506, PhysRevB.92.035153}, the p-wave \cite{PhysRevB.89.014506} and pairings in different superconducting scenarios, leading to unconventional superconducting states, are also of interest \cite{RevModPhys.84.1383}. Since the desired effective theory is essentially fixed by the general requirements of chiral symmetry breaking and charge symmetry breaking, we will construct a specific pairing (intra-node s-wave) that, once condensed, results in an effective theory of a superconductor with dynamical axion interaction. One can expect that the phenomenological features of this model, such as the penetration length to be discussed later, are shared with any model that displays the same symmetries and symmetry breaking patterns.

Considering the formation of condensates that breaks charge symmetry as well as chiral symmetry,  one expects the system to be characterized by four active degrees of freedom (two charges and two chiralities).A simple choice is to encode those degrees of freedom in two complex fields that represent two possible condensates. 
\begin{eqnarray}
\label{orderparameters}
\Phi_R(x) &=& \lambda_R (\bar{\psi}_c P_R \psi) = \lambda_R \psi_R(x)\psi_R(x)\\
\Phi_L(x)  &=& \lambda_L (\bar{\psi}_c P_L \psi) = \lambda_L \psi_L(x)\psi_L(x)
\end{eqnarray}
Where $\lambda_R$ and $\lambda_L$ are couplings of mass dimension $-1$ and
$\psi_c =\smqty(\sigma_2 \psi^{\dagger}_R\\-\sigma_2 \psi^{\dagger}_L)$
 is the charge conjugate spinor field. Note that $\Phi_R(x)$ and $\Phi_L(x)$ carry the same charges ($2e$ if $e$ is the fermion charge) but have opposite chirality. 

The condensation of these operators is supposed to be implied by the four fermions interactions    
\begin{eqnarray}
\label{pairingsSC}
\lambda^2_R (\bar{\psi}_c P_R \psi) (\bar{\psi} P_L \psi_c) + \lambda^2_L (\bar{\psi}_c P_L \psi) (\bar{\psi} P_R \psi_c) = \lambda^2_R \psi_R\psi_R\psi^{\dagger}_R\psi^{\dagger}_R + \lambda^2_L \psi_L\psi_L\psi^{\dagger}_L\psi^{\dagger}_L
\end{eqnarray}
It is a dynamical question whether these couplings are able to give rise to the condensates. If this happens the system will develop a superconducting phase once $\Phi_R(x)$ and $\Phi_L(x)$ are charged. The fermionic action can be written as 
\begin{align}
	\label{diracactiondoubling}
	S = \int \dd[4]{x}  \frac{1}{2}\bar{\Psi}(x) &\left[ \left(i\slashed{\partial} + \slashed{b}\gamma^5 + ie\slashed{A}(x) \rho_3\right)  + \left(\lambda_R\Phi_L^{\ast}  P_R + \lambda_L\Phi_R^{\ast}  P_L \right) (\rho_1 - i \rho_2)\right.\nonumber\\
	&\left. + \left(\lambda_R\Phi_L  P_R + \lambda_L \Phi_R  P_L \right)  (\rho_1 + i \rho_2) \right]\Psi(x),
\end{align}	
Where we define the enlarged spinor $\Psi=\smqty(\psi\\\psi_c)$ and $\bar{\Psi}=\smqty(\bar{\psi} & \bar{\psi}_c)$ and also added  another layer of matrix structure, the Pauli matrices $\rho$, acting on ``charge space''. Thus, the total matrix structure schematically is
\begin{eqnarray}
	\label{matrixstructure}
	\Gamma = \sigma \otimes \tau \otimes \rho
\end{eqnarray} 
with $\sigma$, $\tau$, and $\rho$ acting on the spin, handiness, and charge, respectively. In this notation, the relevant matrices are given by
\begin{subequations}\label{matrixstructure2}
	\begin{align}
	\gamma^{0} &\rightarrow  \sigma_0 \otimes \tau_1 \otimes \rho_0\\
	\gamma^{i} &\rightarrow  i\sigma_i \otimes \tau_2 \otimes \rho_0\\
	\gamma^{5} &\rightarrow  -\sigma_0 \otimes \tau_3 \otimes \rho_0
	\end{align}
\end{subequations}
In matrix notation in $\rho$ space the action is 
\begin{eqnarray}
\label{diracactiondoublingmatrix}
S = \int \dd[4]{x}  \frac{1}{2}\bar{\Psi}(x) \left(\begin{array}{cc}
i\slashed{\partial} + \slashed{b}\gamma^5 + ie\slashed{A}(x) & {\lambda_R\Phi_L  P_R + \lambda_L \Phi_R  P_L} \\
\lambda_R {\Phi^{\ast}_L  P_R + \lambda_L \Phi_R^{\ast}  P_L } & i\slashed{\partial} + \slashed{b}\gamma^5 -  ie\slashed{A}(x)
\end{array}\right) \Psi(x),
\end{eqnarray}	

The system may be characterized by the following transformations:
\begin{itemize}
	
	\item $U(1)$ gauge symmetry 
	\begin{subequations}\label{u1gauge}
		\begin{align}
			\Psi(x) &\rightarrow e^{-i\alpha(x)\rho_3}\Psi(x)\\
			A_{\mu} &\rightarrow  A_{\mu} - \frac{i}{e}\partial_{\mu}\alpha(x)\\
			\Phi_{R/L}(x) &\rightarrow e^{-i2\alpha(x)}\Phi_{R/L}(x)
		\end{align}
	\end{subequations}

	\item $U(1)$ (global) chiral symmetry (anomalous)
	\begin{subequations}\label{u1chiral}
		\begin{align}
			\Psi(x) &\rightarrow e^{-i\beta\gamma^5}\Psi(x)\\
			\Phi_L(x) &\rightarrow e^{-i2\beta}\Phi_L(x)\\
			\Phi_R(x) &\rightarrow e^{i2\beta}\Phi_R(x)
		\end{align}
	\end{subequations}

	\item Charge conjugation ($C$)
	\begin{subequations}\label{chargeconj}
		\begin{align}
			\Psi(x) &\rightarrow \rho_{1}\Psi(x)\\
			A_{\mu} &\rightarrow  -A_{\mu}\\
			\Phi_{R/L}(x) &\rightarrow \Phi_{L/R}^{\ast}(x)
		\end{align}
	\end{subequations}

	\item Parity ($P$): $Px = (t,-x,-y,-z)$
	\begin{subequations}\label{parity}
		\begin{align}
			\Psi(x) &\rightarrow i \tau_{1}\Psi(Px)\\
			A_{\mu}(x) &\rightarrow  \qty(A_{0}(Px),-A_i(Px))\\
			\Phi_{R/L}(x) &\rightarrow \Phi_{L/R} (Px)
		\end{align}
	\end{subequations}

	\item Time reversal  (T): $Tx = (-t,x,y,z)$
	\begin{subequations}\label{timerev}
		\begin{align}
			\Psi(x) &\rightarrow -\sigma_{2}\Psi(Tx)\\
			A_{\mu}(x) &\rightarrow  \qty(A_{0}(Tx),-A_{i}(Tx))\\
			\Phi_{R/L}(x) &\rightarrow -\Phi_{R/L}(Tx)
		\end{align}
	\end{subequations}

\end{itemize} 
The term $\bar{\Psi}\slashed{b}\gamma^5 \Psi$, where $b_{\mu}$ is a background vector, explicitly breaks $P$ ($T$) if $b_{\mu}$ is time-like (space-like). In the main part of this paper we will consider only the case $b_\mu=0$, but in this section we keep it for completeness.

Upon condensation we have
\begin{eqnarray}
	\label{orderparameters2}
	\langle \Phi_R(x) \rangle &=& \lambda_R v^3_R e^{i\delta_R}\\
	\langle \Phi_L(x) \rangle   &=& \lambda_L v^3_L e^{i\delta_L}
\end{eqnarray}
Any choice of parameters breaks $T$ once the system undergoes condensation. If $\lambda_R v^3_R= \lambda_L v^3_L$ then we have the following choices:
\begin{itemize}
	\item $\delta_R=\delta_L$, $P$ is preserved and $C$ is broken;
	\item $\delta_R=-\delta_L$, $C$ is preserved and $P$ is broken;
	\item $\delta_R=\delta_L = 0$ then $C$ and $P$ are preserved.
\end{itemize}
The chiral symmetry is anomalous, which means that it is not a true symmetry of the theory, and the gauge redundancy undergoes a Higgs mechanism. The effective action can be constructed by considering fluctuations of the phases around the vacuum values  $\delta_R $ and $\delta_L$
\begin{eqnarray}
	\label{orderparameters3}
	\Phi_R(x) &=& \lambda_R v^3_R e^{i\frac{\phi_R(x)}{f_R}}\\
	\Phi_L(x)  &=& \lambda_L v^3_L e^{i\frac{\phi_L(x)}{f_L}}
\end{eqnarray}
where $\phi_R(x)$ and $\phi_L(x)$ are the fluctuations. We also perform the redefinition
\begin{eqnarray}
	\label{chargechiraltransform}
	\Psi(x) \rightarrow  e^{i \frac{1}{4}\left(\frac{\phi_R(x)}{f_R} - \frac{\phi_L(x)}{f_L}\right)\gamma^5 }\Psi(x) ,
\end{eqnarray}	
Taking into account the non-trivial Jacobian of the fermionic measure and considering for simplicity $\lambda_R = \lambda_L = \lambda$ and  $v_R= v_L =  v$ we obtain
\begin{align}
	\label{diracAxion2}
	S = \frac{e^2}{16\pi^2} \int \dd[4]{x} &\frac{1}{4}\left(\frac{\phi_R(x)}{f_R} - \frac{\phi_L(x)}{f_L}\right) \varepsilon^{\mu\nu\rho\sigma}F_{\mu\nu}F_{\rho\sigma} \nonumber\\
	+\int \dd[4]{x}  &\frac{1}{2}\bar{\Psi}(x) \left(i\slashed{\partial} + \slashed{b}\gamma^5 - \frac{1}{4}\slashed{\partial} \left(\frac{\phi_R(x)}{f_R} - \frac{\phi_L(x)}{f_L}\right) \gamma^5  + ie\slashed{A}(x) \rho_3 \right)\Psi(x) \nonumber\\
	+ \int \dd[4]{x} & \frac{\lambda^2 v^3}{2} \bar{\Psi}(x) \left(  e^{-i \frac{1}{2}\left(\frac{\phi_R(x)}{f_R} + \frac{\phi_L(x)}{f_L}\right)}   (\rho_1 - i \rho_2)  +e^{i \frac{1}{2}\left(\frac{\phi_R(x)}{f_R} + \frac{\phi_L(x)}{f_L}\right)}   (\rho_1 + i \rho_2)  \right)\Psi(x),
\end{align}
Performing now yet another redefinition  
\begin{eqnarray}
	\label{chargechiraltransform2}
	\Psi(x) \rightarrow  e^{i \frac{1}{4}\left(\frac{\phi_R(x)}{f_R} + \frac{\phi_L(x)}{f_L}\right)\rho_3 }\Psi(x) ,
\end{eqnarray}	
the action becomes 
\begin{align}
	\label{diracAxion3}
	S = \frac{e^2}{16\pi^2} \int \dd[4]{x} &\frac{1}{4}\left(\frac{\phi_R(x)}{f_R} - \frac{\phi_L(x)}{f_L}\right) \varepsilon^{\mu\nu\rho\sigma}F_{\mu\nu}F_{\rho\sigma} \nonumber\\
	+\int \dd[4]{x}  &\frac{1}{2}\bar{\Psi}(x) \left(i\slashed{\partial} + \slashed{b}\gamma^5 - \frac{1}{4}\slashed{\partial} \left(\frac{\phi_R(x)}{f_R} - \frac{\phi_L(x)}{f_L}\right) \gamma^5  \right. \nonumber\\
	&\left.  + ie\gamma^{\mu}\left( A_{\mu}(x)  + i \frac{1}{4e}\partial_{\mu}\left(\frac{\phi_R(x)}{f_R} + \frac{\phi_L(x)}{f_L}\right) \right)\rho_3  + 2\lambda^2 v^3 \rho_1\right)\Psi(x) 
\end{align}
or
\begin{align}
	\label{diracAxion4}
	S = \frac{e^2}{32\pi^2} \int \dd[4]{x} &\left(\frac{\theta(x)}{f} + \theta_0(x)\right) \varepsilon^{\mu\nu\rho\sigma}F_{\mu\nu}F_{\rho\sigma} \nonumber\\
	+\int \dd[4]{x}  &\frac{1}{2}\bar{\Psi}(x) \left(i\slashed{\partial}  - \frac{1}{2f}\slashed{\partial} \theta(x) \gamma^5  + ie\gamma^{\mu}\left( A_{\mu}(x)  + i \frac{1}{2ef'}\partial_{\mu}\theta' \right)\rho_3  + 2\lambda^2 v^3 \rho_1\right)\Psi(x) 
\end{align}
where we defined $\frac{\theta(x)}{f} + \theta_0(x) = \frac{1}{2}\qty(\frac{\phi_R(x)}{f_R} - \frac{\phi_L(x)}{f_L})$  and $\frac{\theta'(x)}{f'} = \frac{1}{2}\qty(\frac{\phi_R(x)}{f_R} + \frac{\phi_L(x)}{f_L})$, with  $\theta_0(x) = 2b_{\mu}x^{\mu}$. Note that $\theta'(x)$ is the would-be Goldstone boson that is combined with the gauge field in the Higgs mechanism to furnish the gauge invariant piece $A_{\mu}(x)  + i \frac{1}{2ef'}\partial_{\mu}\theta'$ representing a longitudinal term for the vector field, thus leading to a consistent mass term for the photon, characterizing the Meissner effect. We also note that a mass term for the field $\theta(x)$ will be induced non-perturbatively due to the fermionic condensate, explicitly
\begin{align}
	\label{condtheta}
	\langle \bar{\Psi}(x) \rho_1 \Psi(x) \rangle = 4v^3\cos[2](\frac{\theta}{f}+\theta_0)
\end{align}
In the previous derivation, since we have ignored the compactness of the fields $\theta (x)$ and $\theta' (x)$, we are not considering the contribution of singular states such as vortices that can be described by multivalued fields \cite{Braga_2020}. These non-perturbative effects are indispensable if one is interested in a comprehensive characterization of the system. In the present case, the vortices associated with $\theta' (x)$ are the usual ones from a superconductor and carry quantized magnetic flux. The vortices of $\theta (x)$  are more interesting and were called chiral vortices  in \cite{PhysRevB.87.134519}. They don’t carry magnetic flux but are responsible for a non-conservation of the naive supercurrent of the superconductor \cite{Braga_2016}, see also \cite{PhysRevB.93.174501}. Both kinds of vortices must be taken into account if one is interested in the topological features of the superconducting state and, in fact, one can construct the corresponding effective topological field theories by reasoning about the dilution and condensation of such configurations \cite{Braga_2016}. However, since our goal is the perturbative analysis of the resulting effective theory, such non-perturbative effects are not relevant.

Computing the fermionic field integration (where the fermionic determinant may be evaluated as a derivative expansion of gauge-invariant terms) and taking into account the non-perturbative mass term leads us to the general form for the electromagnetic response of the system
\begin{align}\label{axion-proca}
S_{MP} =\int \dd[4]{x}
\left(
-\frac{1}{4}F_{\mu\nu}F^{\mu\nu}+\frac{1}{2}M^{2}A_{\mu}A^{\mu}+\frac{1}{2}\partial_\mu \theta \partial^\mu \theta -\frac{1}{2}m^{2}\theta^{2}+\frac{1}{4}g\left(\theta + \frac{\theta_0}{g}\right) \tilde{F}_{\mu\nu}F^{\mu\nu}
\right) 
\end{align}
where $g\sim \frac{1}{f} \sim \frac{1}{\lambda^2 v^3}$ and $M\sim \lambda^2 v^3$, $m\sim \frac{\lambda v^3}{f} \sim \frac{1}{\lambda}$. These relations make contact with the microscopic theory we have been developing in this section. They set the scaling behavior of the parameters of the effective theory as function of the ones of the microscopic theory. In what follows we will not adhere to these relations and instead consider, for the sake of computations, $M$, $m$ and $g$ as independent quantities. However, later in this paper we will comment on the relations with the microscopic theory. 

The effective action \ref{axion-proca} describes the electromagnetic response of a microscopic system characterized by chiral and charge condensates, whose fluctuations give rise to the dynamic of the axion field and to the photon mass, through the Higgs mechanism. The same effective theory can be obtained by dimension reduction from a $5D$ theory \cite{PhysRevB.87.134519} and also from general reasoning about condensation of charges and defects guided by symmetry considerations \cite{Braga_2016}. But it is important to point out that we arrived at this action considering an interaction that makes contact with usual superconducting couplings in doped Weyl metals \cite{PhysRevB.92.035153}. This goes back to our initial considerations regarding the possible pairings. From the point of view of the resulting effective theory (that can originate from different pairings, i.e. instability in a Weyl semimetal system) the main point is that if one is interested in the identification of the relevant low energy degrees of freedom, including possible defects, and the ensuing non-trivial topological features of the superconducting states, the answer seems to involve topological BF theories, as discussed in \cite{Hansson_2004} for the usual superconductor, in \cite{Hansson_2015} for a p-type superconductor and in \cite{Braga_2016} for the axionic superconductor. The different possible pairings, in this case, will enter the analysis because they are responsible for defining the low energy degrees of freedom that are relevant for the topological description of the system. But, if one is interested in the general features of the electromagnetic response, as we are in the present work, the answer has less freedom and is essentially fixed by symmetry with the microscopic theory furnishing the parameters of the effective theory, as discussed above.

Our task now in the next sections is to compute the modifications on the Yukawa potential and, via the analysis of quantum corrections to the London's length, the Meissner effect induced by axionic fluctuations. For the present work, we will set $\theta_0=0$, which will simplify, considerably, the computations, meaning that we will be analyzing the physics of a Dirac semimetal ($b_{\mu} =0$).

%----------------------------------------------------------------------------------------
\section{Axion-Proca electrodynamics}
The model is defined by the following action (in natural units and $\text{diag}(g_{\mu\nu})=(1,-1,-1,-1)$)
\begin{align}\label{initial.action}
\tilde{S}_{\overline{g}}=\int d^{4}x
\left(
-\frac{1}{4}f_{\mu\nu}f^{\mu\nu}+\frac{1}{2}\overline{M}^{2}\overline{a}_{\mu}\overline{a}^{\mu}+\frac{1}{2}\partial_\mu \overline{\theta}\partial^\mu \overline{\theta}-\frac{1}{2}\overline{m}^{2}\overline{\theta}^{2}+\frac{1}{4}\overline{g}\overline{\theta}\tilde{f}_{\mu\nu}f^{\mu\nu}
\right) 
\end{align}

This effective action describes the dynamics of a massive vector field (Proca) $\overline{a}_\mu(x)$ and a massive pseudo-scalar field $\overline{\theta}(x)$, displaying an axion-like interaction. Envisaging the renormalization analysis to follow, the field strength tensor is written in terms of ``bare'' quantities, so $f_{\mu\nu}=\partial_\mu \overline{a}_\nu -\partial_\nu \overline{a}_\mu$, with the dual tensor $\tilde{f}_{\mu\nu}=\dfrac{1}{2}\epsilon_{\mu\nu\sigma\rho}f^{\sigma\rho}$. The coupling constant $\overline{g}$ has a mass of dimension $-1$, so power counting indicates that this theory is {\it nonrenormalizable}. This Lagrangian must be understood as describing the physics at energies much lower than the cut-off $\Lambda_{UV} \sim 1/\overline{g}$.

Since we will focus on the computation of the vector field propagator up to 1-loop order, some particular simplifications can be model using symmetry characteristics. For example, the lack of gauge invariance allows for terms like $M^2_1(\overline{a}^2)^2$ to be included at order $\overline{g}^2$, but an odd number of $\overline{a}_\mu$ will not contribute because this would break the discrete symmetry $\overline{a}_\mu (x)\to -\overline{a}_\mu (x)$. The same does not apply to the case for the scalar field because the coupling does have an odd number of $\overline{\theta}$'s. One contemplate possibility is $(\overline{a}^2)^3$, but such a term will give a six photon vertex that is only relevant to the propagator if taken at 2-loops. One algorithm that describes a similar process, for Proca-electrodynamics, can be found in \cite{GallegoCadavid:2019zke}. Lastly, the most general contribution must include terms composed with the dual field strength $\tilde{f}_{\mu\nu}$ but, since we are only interested in the contribution to the massive photon two-point function, they will be zero after we impose momentum conservation at the vertex.

All workable terms of order $\overline{g}^2$ can be organized in three new Lagrangian pieces
\begin{align}
	\mathcal{{\overline L}}_{\theta g^2}&=-\frac{1}{2}\overline \theta^2 m_1^2+\frac{1}{2} \overline C_\theta(\partial \overline \theta)^2+\frac{1}{2\overline m_s^2}(\partial_\mu \overline \theta)\square(\partial^\mu \overline \theta)\\
	\mathcal{{\overline L}}_{a g^2}&=\frac{1}{2}\overline M_1^2 \overline a^2-\frac{1}{4}\overline C_{f}f^2+\frac{1}{2 \overline m_{gh}^2}(\partial f)^2+\frac{1}{4!}\frac{1}{2}\overline a^4 \overline C_{4}-\frac{1}{4!}\frac{1}{4}\frac{\overline a^2}{\overline M_2^2} f^2\\
	\mathcal{{\overline L}}_{a\theta g^2}&=-\frac{1}{2}\overline C_{a\theta} \overline \theta^2 \overline a^2 +\frac{1}{4}\frac{\overline \theta^2}{\overline m_{\theta f}^2} f^2
	\label{g2pertub}
\end{align}
These modifications can be divided further into two groups by noticing that some terms can be absorbed in parameter redefinitions in the process of normalization since they are of order $g^2$. The other terms with higher derivatives (i.e. $(\partial^2 f)$ and $(\partial_\mu \overline{\theta})\square(\partial^\mu \overline{\theta})$), will generate ghost contributions to the free field propagator. Nevertheless, in this model, it is possible to eliminate this kind of non-physical contribution performing field redefinitions so that the free propagator will remain well behaved and unitary. Most of the discussion is based on \cite{Villalba-Chavez:2018eql} and \cite{Grinstein:2007mp, Accioly:2005xf}, and the mathematical detail for our case that deviate from those works are described in appendix \ref{bare.to.renor.ghost.elimination}. The Lagrangian with redefined parameters reads
\begin{align}
	\mathcal{L}_R&= -\frac{1}{4}Z_3F^2+\frac{1}{2}M^2 Z_M Z_3 A^2+\frac{1}{2}Z_\theta(\partial \theta)^2-\frac{1}{2}Z_m Z_\theta m^2\theta^2\nonumber\\
&\quad+\frac{Z_g}{4}g\theta \tilde{F}^{\mu\nu}F_{\mu\nu}+\frac{\delta_{s}}{2 m_s^2}(\partial_\mu  \theta)\square(\partial^\mu  \theta)+\frac{\delta_{gh}}{2  m_{gh}^2}(\partial F)^2
+\mathcal{L}_{4\gamma}+\mathcal{L}_{2\gamma,2\theta}
\end{align}
with the new interaction terms
\begin{align}
	\mathcal{L}_{4\gamma}=\frac{1}{4!}\frac{1}{2}Z_{4}C_{4} A^4-\frac{1}{4!}\frac{Z_5}{4}\frac{ A^2}{ M_2^2} F^2\qquad\&\qquad
	\mathcal{L}_{2\gamma,2\theta}=-\frac{1}{2}Z_{a \theta} C_{a\theta}  \theta^2  A^2 +\frac{1}{4}Z_{\theta f}\frac{ \theta^2}{ m_{\theta f}^2} F^2
\end{align}
of order $g^2$ (since $C_{4},C_{a\theta},M_2^{-2}$ and $m_{\theta f}^{-2}\in\order{g^2}$). All these interactions will furnish 1-loop contributions to the massive vector self-energy.
%----------------------------------------------------------------------------------------
%	Self energy graph
% results of sec 1.4 from the mathematica file
%----------------------------------------------------------------------------------------
\section{Photon self-energy}
\begin{figure}[hbtp]
	\centering
	\includegraphics[width=0.3\textwidth]{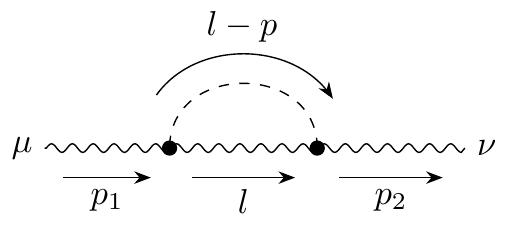}
	\hspace{0.5cm}
	\includegraphics[width=0.3\textwidth]{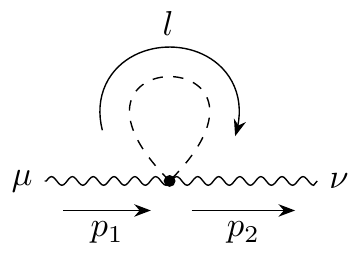}
	\hspace{0.5cm}
	\includegraphics[width=0.3\textwidth]{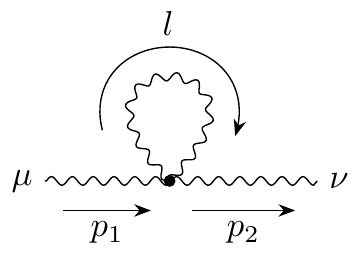}
	\caption{Sum of Feynman's graphs that contribute to the photon self energy in axion-Proca electrodynamics. In order from left to right: \textit{axion loop} $K^{(1)}_{\mu\nu}$, \textit{photon-axion loop} $K^{(2)}_{\mu\nu}$ and \textit{photon-photon loop} $K^{(3)}_{\mu\nu}$}
	\label{graph.feynman.sum}
\end{figure}

We want to compute quantum corrections to the massive vector self-energy introduced by axion fluctuations. The dressed massive vector propagator will include 1-loop contributions that originates from the axion coupling ($\order{g}$) and from $\mathcal{L}_{4\gamma}$ and $\mathcal{L}_{2\gamma,2\theta}$ ($\order{g^2}$). The exact Green function for the photon $G^{\mu\nu}(p)$ is given by the geometric sum of 1PI graphs
\begin{align}
i G^{\mu\nu}(p)&=
\begin{tikzpicture}
  \begin{feynman}[small, layered layout]
    \vertex (a);
    \vertex [right=of a] (b);
    \diagram* {
      (a) -- [photon] (b),
    };
  \end{feynman}
\end{tikzpicture}
+
  \begin{tikzpicture}
  \begin{feynman}[small, layered layout]
    \vertex (a);
    \vertex [right=of a] (b);
	\vertex [right=of b] (c);
    \vertex [right=of c] (d);
    \diagram* {
      (a) -- [photon] (b),
      (b) -- [photon] (c),
      (c) -- [scalar, half right] (b),
      (c) -- [photon] (d),
    };
  \end{feynman}
  \filldraw [black] (b) circle (1.5pt);
  \filldraw [black] (c) circle (1.5pt);
\end{tikzpicture}
+
\begin{tikzpicture}
  \begin{feynman}[small, layered layout]
    \vertex (a);
    \vertex [right=of a] (b);
    \vertex [above=of b] (b1);
	\vertex [right=of b] (c);	
    \diagram* {
      (a) -- [photon] (b),
      (b) -- [scalar,out=135, in=45, loop, min distance=1cm] b,
      (b) -- [photon] (c),
    };
  \end{feynman}
  \filldraw [black] (b) circle (1.5pt);
\end{tikzpicture}
+
\begin{tikzpicture}
  \begin{feynman}[small, layered layout]
    \vertex (a);
    \vertex [right=of a] (b);
    \vertex [above=of b] (b1);
	\vertex [right=of b] (c);	
    \diagram* {
      (a) -- [photon] (b),
      (b) -- [photon,out=135, in=45, loop, min distance=1cm] b,
      (b) -- [photon] (c),
    };
  \end{feynman}
  \filldraw [black] (b) circle (1.5pt);
\end{tikzpicture}
+\cdots\\
&=i G_0^{\ \mu\nu}(p)+i G_0^{\ \mu\sigma}(p)\left(i \Pi_{\sigma\rho}(p)\right)i G_0^{\ \rho\nu}(p)+\order{g^4}\label{exact.Green.function}
\end{align}
where $G_0^{\mu\nu}(p)$ is the free massive vector propagator, defined as $G_{0}^{\mu\nu}(p)=-iP^{\mu\nu}(p)/(p^2-M^2)$ with $P_{\mu\nu}(p)=g_{\mu\nu}-p_\mu p_\nu /M^2$, and $i \Pi_{\sigma\rho}(p)$ is the 1-loop contributions (consult figure \ref{graph.feynman.sum} for exact Feynman's diagram anatomy) with the additional factors given by counterterms.
\begin{align}\label{eq:Pi.full}
	i \Pi_{\sigma\rho}(p)=\sum_i^{3} K_{\sigma\rho}^{(i)}(p^2)-i(Z_3-1)(p^2g_{\sigma\rho}-p_\sigma p_\rho)+i(Z_M-1)(Z_3-1)M^2 g_{\sigma\rho}+i\frac{\delta_{gh}}{m_{gh}^2}p^2(p^2g_{\sigma\rho}-p_\sigma p_\rho)
\end{align}
%----------------------------------------------------------------------------------------
%	Loop integral
%----------------------------------------------------------------------------------------
\subsection{Loop integral}
The axion coupling introduces a momentum dependent vertex that can be written schematically as $ V_{\mu\nu}(p_1,p_2)=(igZ_g)\epsilon_{\mu\nu\alpha\beta}p_1^\alpha p_2^\beta$ where the vector line carries ingoing momentum $p_1$ and outgoing momentum $p_2$. The vertex construction results in
\begin{align}
	K_{\mu\nu}^{(1)}=\int\frac{\dd[4]{l}}{(2\pi)^4}V_{\mu\sigma}(-p_1,l)G_{0}^{\sigma\rho}(l)\Delta_{0}(l-p)V_{\rho\nu}(l,-p_2)
\end{align}
where $\Delta_{0}(p)=i/(p^2-m^2)$ is the free massive pseudo scalar propagator in momentum space. Since $Z_g =1+O[g^2]$ (so that $(igZ_g)^2\sim -g^2$) we can write the contribution from the \textit{axion loop} graph as
\begin{align}
	K_{\mu\nu}^{(1)}(p^2)&=-g^2 \int\frac{\dd[4]{l}}{(2\pi)^4}\frac{Y_{\mu\nu}(p,l)}{l^2-M^2}\frac{1}{(l-p)^2-m^2}
\end{align}
with $Y^{\mu\nu}(p,l)=g^{\mu  \nu } \left(l^2 p^2-(l\cdot p)^2\right)+l^{\mu } \left(p^{\nu } (l\cdot p )-p^2 l^{\nu }\right)+p^{\mu } \left(l^{\nu } (l\cdot p )-l^2 p^{\nu }\right)$. Using the standard Feynman parametrization, the expression becomes
\begin{align}
	K_{\mu\nu}^{(1)}(p^2)&=-\frac{g^2}{2}(g_{\mu\nu}p^2-p_\mu p_\nu)\int\frac{\dd[4]{q}}{(2\pi)^4}\int_0^1 \dd{s} \frac{q^2}{(q^2-\Delta(s,p^2))^2}
\end{align}
with $\Delta(s,p^2)=m^2 s-M^2(s-1)-p^2 s(1-s)$. Even though gauge invariance is explicitly broken by the mass term, the longitudinal component is effectively decoupled and the result can still be written using the usual transverse operator 
\begin{align}
	K_{\mu\nu}^{(1)}(p^2)&=(g_{\mu\nu}p^2-p_\mu p_\nu)k^{(1)}(p^2).
\end{align}
This can be formally established by a Ward identity (\cite{vanHees:2003dk}) showing that only the transverse part will contribute to the final result. Now we must extend $k^{(1)}(p^2)$ to $D-$dimensions and redefine the dimensional coupling as $g\to g\mu^{\frac{4-D}{2}}$  ($\mu$ is an arbitrary parameter of mass dimension 1 so that the coupling $g$ is now dimensionless). Also, this rescaling must be followed by a redefinition of the Wilson parameters $b_i\to b_i \mu^{\frac{D-4}{2}} $ so that $b_i$ is also dimensionless. Integrating over $q$ and expanding for $D=4-\epsilon$ with $\epsilon\to0$ we obtain
\begin{align}\label{eq:k1}
	k^{(1)}(p^2)&=-\frac{ig^2}{16\pi^2}
	\qty[
	\frac{2}{\epsilon}\left(\frac{m^2}{2}+\frac{M^2}{2}-\frac{p^2}{6}\right)
	-\int_0^1 \dd{s}
		\Delta\log\frac{\Delta}{\tilde{\mu}^2}
	]
\end{align}
with the usual definition $\tilde{\mu}^2=e^{-\gamma}4\pi\mu^2$ ($\gamma$ is the Euler-Mascheroni constant). In this computation, any part that is not divergent or that don't have any kind of discontinuity can be ignored since they will simply be absorbed by a finite redefinition of the original action.

The same process is used to compute $K_{\mu\nu}^{(2,3)}$ resulting in
\begin{align}\label{eq:k2.e.2}
	k^{(2)}(p^2)+k^{(3)}(p^2)=\frac{i}{16\pi^2}
	\qty[
	m^2\qty(\frac{2}{\epsilon}-\log(\frac{m^2}{\tilde{\mu}^2}))\left(
		C_{a\theta}+\frac{p^2}{ m_{\theta f}^2}
	\right)
	-
	M^2\qty(\frac{2}{\epsilon}-\log(\frac{M^2}{\tilde{\mu}^2}))\left(
	3C_4+\frac{p^2+M^2}{2 M_2^2}
	\right)
	]
\end{align}
%----------------------------------------------------------------------------------------
%	Renormalization
%----------------------------------------------------------------------------------------
\subsection{Renormalization}
\label{sec:renormalization}
%----------------------------------------------------------------------------------------
Using equations (\ref{eq:k1}),(\ref{eq:k2.e.2}) in (\ref{eq:Pi.full}) results in
\begin{align}
	i \Pi_{\mu\nu}(p)&=i\Pi(p^2)g_{\mu\nu}+(p_\mu p_\nu-\text{terms})\\
	\Pi(p^2)&=\frac{1}{16\pi^2}\qty[\Pi^{(0)}+p^2 \Pi^{(2)}(p^2)+p^4 \Pi^{(4)}-p^2\delta_3 +(\delta_M+\delta_3)M^2+p^4\frac{\delta_{gh}}{m^2_{gh}}]\label{PI}
\end{align}
The exact Green's function at one loop, in this context, is given by
\begin{align}\label{exact.green.fun}
i G_{\mu\nu}(p^2)=-i\frac{g_{\mu\nu}}{p^2(1+\Pi^{(2)})-(M^2-\Pi^{(0)})+p^4\Pi^{(4)}(p^2)}+(p_\mu p_\nu-\text{terms})
\end{align}
with
\begin{align}
\Pi^{(0)}&=
\qty(\frac{2}{\epsilon}-\log(\frac{m^2}{\tilde{\mu}^2}))m^2C_{a\theta}
-
M^2\qty(\frac{2}{\epsilon}-\log(\frac{M^2}{\tilde{\mu}^2}))\left(
3C_4+\frac{M^2}{2 M_2^2}
\right)+M^2\delta_{M}\label{Pi.0}\\
\Pi^{(2)}&=
g^2\left(
\frac{2}{\epsilon}\left(\frac{m^2}{2}+\frac{M}{2}\right)
+\int_0^1 \dd{s} \Delta \log \frac{\Delta}{\tilde{\mu}^2}		
\right)
+
\qty(\frac{2}{\epsilon}-\log(\frac{m^2}{\tilde{\mu}^2}))
\frac{m^2}{ m_{\theta f}^2}
-
\qty(\frac{2}{\epsilon}-\log(\frac{M^2}{\tilde{\mu}^2}))
\frac{M^2}{2 M_2^2}\nonumber\\
&\qquad+\frac{\delta_3}{p^2}\qty(M^2-p^2)\label{Pi.2}\\
\Pi^{(4)}&=g^2 \frac{2}{\epsilon}\frac{1}{6}+\frac{\delta_{gh}}{m^2_{gh}}\label{Pi.4}
\end{align}
This expression is correct up to $\order{g^4}$ (with the exception of $\frac{\delta_{gh}}{m^2_{gh}}$) and any finite term\footnote{All $\delta$ ware redefined to include the $16\pi^2$ factor}. 

Before proceeding with the renormalization process, it should be clear that this expression results in the one found in \cite{Villalba-Chavez:2018eql} once we set $M^2=0$. As a consequence of the restored gauge invariance, no term $\sim\Pi^{(0)}$ can be found (note that $C_{a \theta}$ would not be included in $\mathcal{{\overline L}}_{a\theta g^2}$ in \eqref{g2pertub} from the beginning).

We would like to draw attention to a characteristic of our model regarding the subtraction scheme choice, but first, it is interesting to comment on the potential felt by a test charge in the massless photon limit. Axion fluctuations are responsible for a correction of the Coulomb electrostatic potential, felt by a test charge $e$, that can be written as
\begin{align}
\tilde{V}_{M=0}(p^2)=\frac{e^2}{p^2}
\qty[
1+\frac{1}{p^2}
\left(
\Pi^{(2)}(p^2_0)-\Pi^{(2)}(p^2)+(p_0^2-p^2)\Pi^{(4)}
\right)
]
+O[e^2 g^4]
\end{align}
in momentum space evaluated at $p$ concerning it's value at the scale $p_0$. Note that $\Pi^{(4)}$ is constant at this order and can be set to zero by imposing $\overline{MS}$ scheme. It is then physically sensible to make contact with the measured electric charge by defining the potential to have the Coulomb form at spatial infinity, or equivalently at $p_0=0$, where the axion effect should be negligible. That is, to fix $p_0$ is sufficient to impose that the potential is of the usual Coulomb type at $p_0=0$ resulting in $e$ being the observable electric charge. This works as a renormalization condition fixing the ambiguity in $\Pi^{(2)}$.

The electrostatic potential felt by a test charge in this massive photon setting can be written as
\begin{align}
\tilde{V}(p^2)=\frac{e_R^2}{p^2-M^2}
\qty(1+\frac{1}{p^2-M^2}
\left(
\frac{p^2_0 \Pi^{(2)}(p^2_0)+p^4_0\Pi^{(4)}}{p^2_0-M^2}		
-\frac{p^2 \Pi^{(2)}(p^2)+p^4\Pi^{(4)}}{p^2-M^2}
\right))	
+O[e_R^2 g^4]
\end{align}
Note that here the scale $p_0$ is defined as the scale where the potential is of the Yukawa type. But now one can not use the asymptotic charge to define a physically motivated renormalization condition as done above in the massless case. The potential of a massive photon is null asymptotically as a result of the screening due to the superconductivity. Physically, due to the massive nature of the photon, test charges will feel no force at spatial infinity. This is a setback for the use of the $\overline{MS}$ scheme because there is no simple way to fix the remaining ambiguity. This problem can be avoided if we impose the so-called on-shell ($OS$) conditions.

It is clear from equation (\ref{exact.green.fun}) that it will be necessary three conditions to fix the singular $\epsilon^{-1}$ contributions that are proportional to $p^0$, $p^2$ and $p^4$. They will be
\begin{align}
\Pi(M^2)&=0\\
\eval{\pdv{\Pi(p^2)}{p^2}}_{p^2=M^2}&=0\\
\eval{\frac{\partial^2 \Pi(p^2)}{(\partial p^2)^2}}_{p^2=M^2}&=0
\end{align}
but before we apply these conditions we must make a $\order{g^4}$ modification
\begin{align}
p^4\frac{\delta_{gh}}{m^2_{gh}}\to \frac{1}{2}(p^2-M^2)^2\frac{\delta_{gh}}{m^2_{gh}}
\end{align}
The first two conditions fix the mass pole location and the residue (so that the physical photon mass is $M^2$ with residue $i$). The third cancel any contribution from $\Pi^{(2)}$ by fixing the ghost counter-term.
Now we can impose these restrictions, resulting in a physically consistent potential clear from any infinities and free parameters. The counter terms obtained are
\begin{align}
\delta_M&=
-g^2 \int_0^1 \dd{s} \qty(m^2 s+M^2 (s-1)^2) \log(\frac{m^2 s+M^2 (s-1)^2}{\mu ^2})- \log(\frac{\mu^2}{m^2})\qty(C_{a\theta}\frac{m^2}{M^2}+\frac{m^2}{m_{\theta f}^2})\nonumber\\
&\qquad
+\frac{1}{\epsilon}\qty(-\frac{2 m^2 C_{a\theta}}{M^2}+6 C_4-\frac{1}{3} g^2 \qty(3 m^2+4 M^2)-\frac{2 m^2}{m_{\theta f}^2}+\frac{2 M^2}{M_2^2})+\log(\frac{\mu^2}{M^2})\qty(3 C_4 +\frac{M^2}{M_2^2}) \\
\delta_3&=
-g^2 \int_0^1 \dd{s}\qty(m^2 s+M^2 (s-1) (2 s-1)) \log(\frac{m^2 s+M^2 (s-1)^2}{\mu ^2})\nonumber\\
&\qquad
+\frac{1}{\epsilon}\qty(-\frac{1}{3} g^2 \qty(3 m^2+5 M^2)-\frac{2 m^2}{m_{\theta f}^2}+\frac{M^2}{M_2^2})-\frac{m^2}{m_{\theta f}^2}\log(\frac{\mu^2}{m^2})+\frac{1}{6}\qty(g^2 M^2+\frac{3M^2}{M_2^2}\log(\frac{\mu^2}{M^2}))\\
\delta_{gh}&=-\frac{2}{3}\frac{g^2 m_{gh}^2}{\epsilon}+\frac{1}{3}g^2 m_{gh}^2
-g^2 m_{gh}^2 \int_0^1 \dd{s} \qty(\frac{M^2(s-1)^2 s^2 s}{M^2 (s-1)^2 +m^2 s}+2s(s-1)\log(\frac{M^2(s-1)^2+m^2 s}{\mu^2}))
\end{align}
so that the result is
\begin{align}\label{eq:pot.renor}
\Pi(p^2)&=-\frac{1}{32 \pi^2}g^2
\left(M^2-p^2\right)\int_0^1 \dd{s}\frac{(s-1) s \left(-2 m^2 p^2 s+M^4 (s-1) s+M^2 p^2 \left(-3 s^2+5 s-2\right)\right)}{m^2 s+M^2 (s-1)^2}
\nonumber\\
&\qquad+\frac{1}{16 \pi^2}g^2
p^2\int_0^1 \dd{s}
\Delta(s,p^2)\log(\frac{\Delta(s,p^2)}{m^2 s+M^2(s-1)^2})
\end{align}
with the previous definition $\Delta(s,p^2)=m^2 s-M^2(s-1)-p^2 s(1-s)$. This is our result for the quantum correction using the OS re-normalization scheme.

Note that the $log$ integrand gives rise to an imaginary part when $p^2>(M+m)^2$
\begin{align}\label{Im.Pi}
	\Im{\Pi(p^2)}&=\frac{1}{96\pi^2}\frac{g^2}{p^2}\qty[\qty(p^2-(M-m)^2)\qty(p^2-(M+m)^2)]^{\frac{3}{2}}
\end{align}
marking the threshold for multiparticle production, with the corresponding spectral function proportional to $\Im{\Pi(p^2)}$.

%%----------------------------------------------------------------------------------------
%%	Potential
%%----------------------------------------------------------------------------------------
\section{Potential}
\begin{figure}[hbtp]
	\centering
	\begin{subfigure}[t]{0.45\linewidth}
		\centering
		\includegraphics[width=0.9\textwidth]{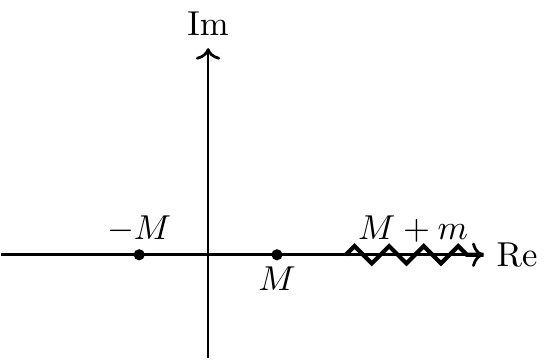}
		\caption{Complex plot of the poles $q=\pm M$ and the cut $q>m+M$.}
		\label{fig.poles}
	\end{subfigure}
	\begin{subfigure}[t]{0.45\linewidth}
		\centering
		\includegraphics[width=.9\textwidth]{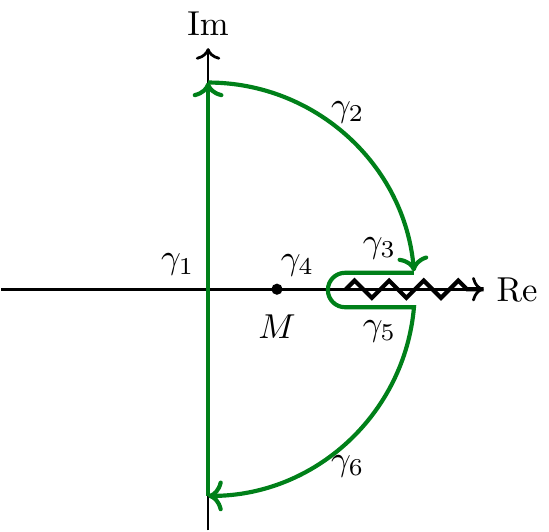}
		\caption{Closed contour know as ``half Pacman".}
		\label{fig.pacman}
	\end{subfigure}
	\caption{Complex plane with $\Re{q}\times\Im{q}$}
\end{figure}

%%----------------------------------------------------------------------------------------
The quantum correction computed in \eqref{eq:pot.renor} allows us to investigate the corresponding correction for electrostatic interaction potential.  The full photon propagator is
\begin{align}
\left<A^\mu (x) A^\nu (y)\right>=\int \frac{\dd[4]{p}}{(2\pi)^4}e^{i p\cdot(x-y)}i G^{\mu\nu}(p)
\end{align}
where $ G^{\mu\nu}(p)$ is the exact propagator, i.e., the propagator for the massive vector field with all its quantum corrections.
Up to 1-loop, we can write
\begin{align}\label{exact.propagator}
G^{\mu\nu}(p)=-i\frac{g^{\mu\nu}}{p^2-M^2}
\left(
1-\frac{\Pi(p^2)}{p^2-M^2}
\right)
+\order{g^4}+(p_\mu p_\nu-\text{terms})
\end{align}
These corrections generate a dressed four potential $\mathcal{A}_\mu (x)$\footnote{This is the same relation used in \cite{Villalba-Chavez:2018eql}. The factor $-i$ follows from the definition of the free propagator (that influences the $i$'s in the exact propagator). Another convention is presented in \cite{Greiner:1992bv}.} given by
\begin{align}
\mathcal{A}_{\mu}(x)=-i\int \frac{\dd[4]{p}}{(2\pi)^4}e^{-i q\cdot x}G_{\mu\nu}(p) \tilde{j}^\nu (p)
\end{align}
Using \ref{exact.propagator} results in
\begin{align}
\mathcal{A}_{\mu}(x)=-\int \frac{\dd[4]{p}}{(2\pi)^4}e^{-i p\cdot x}\frac{\tilde{j}_\mu (p)}{p^2-M^2}
\left(
1-\frac{\Pi(p^2)}{p^2-M^2}
\right)
\end{align}
Now to compute the Yukawa's corrected law we need to use a stationary current $j_\mu(x)$
\begin{align}
j_\mu(x)=e\delta^3(\va{x})\delta_{\mu 0} \to 
\tilde{j}_\mu (p)=2\pi e \delta(p_0) \delta_{\mu 0}
\end{align}
where $e$ is the electric charge, so that\footnote{Remember that $p\cdot x=p^0 x^0 -\va{p}\cdot\va{x}$}
\begin{align}
\mathcal{A}_{0}(\va{x})=e\int \frac{\dd[3]{p}}{(2\pi)^3}e^{i \va{p}\cdot \va{x}}\frac{1}{\abs{\va{p}}^2+M^2}
\left(
1+\frac{\Pi(-\abs{\va{p}}^2)}{\abs{\va{p}}^2+M^2}
\right)
\end{align}
This gives the Fourier transform of the corrected Yukawa potential \cite{Schwartz:2013pla} felt by a negative charge $-e$
\begin{align}
\tilde{V}(\va{p})=-e\tilde{\mathcal{A}}_0 (\va{p})=\frac{-e^2}{\abs{\va{p}}^2+M^2}
\left(
1+\frac{\Pi(-\abs{\va{p}}^2)}{\abs{\va{p}}^2+M^2}
\right)
\end{align}
so that the potential between two identical charges of opposite signs reads
\begin{align}
V(\va{x})=-e^2\int \frac{\dd[3]{p}}{(2\pi)^3}e^{i \va{p}\cdot \va{x}}\frac{1}{\abs{\va{p}}^2+M^2}
\left(
1+\frac{\Pi(-\abs{\va{p}}^2)}{\abs{\va{p}}^2+M^2}
\right)
\end{align}
With this in mind, we can separate this into two contributions
\begin{align}
V_Y(\va{x})=-e^2\int \frac{\dd[3]{p}}{(2\pi)^3}e^{i \va{p}\cdot \va{x}}\frac{1}{\abs{\va{p}}^2+M^2}
\qquad\&\qquad
\delta V_Y(\va{x})=-e^2\int \frac{\dd[3]{p}}{(2\pi)^3}e^{i \va{p}\cdot \va{x}}
\frac{\Pi(-\abs{\va{p}}^2)}{(\abs{\va{p}}^2+M^2)^2}
\end{align}
The computation of the Yukawa potential is well known and results in 
\begin{align}
V_Y(r)=-\frac{e^2}{4\pi}\frac{e^{-M r}}{r}
\end{align}
with $r\equiv\abs{\va{x}}$. To compute $\delta V_{Y}$, we consider the analytic continuation $\abs{\va{p}} \rightarrow i q \in \mathbb{Z}$, which structure is displayed in fig.\ref{fig.poles} (the integrand has a pole at $q =\pm M$ and a cut that starts at $q =(M+m)$).
The complex path, represented in fig.\ref{fig.pacman}, is a ``half-disk" that avoids the branch cut. Here, the integral along $\gamma_1$ is $\delta V_Y(r)$, that after a variable exchange and the identification $q=-i\abs{\va{p}}$ is
\begin{align}
\label{potspacial}
\delta V_Y(r)=\frac{e^2}{4 \pi^2 r i }\int_{-\infty}^{\infty} \dd{q} e^{- r q}\frac{q\Pi(q^2)}{(q^2-M^2)^2}
\end{align}
, and a jump of the cut that can be represented by
\begin{align}
	\Pi(q^2+i\epsilon)-\Pi(q^2-i\epsilon)=\Pi(q^2+i\epsilon)-\Pi(q^2+i\epsilon)^*=2 i \Im{\Pi(q^2+i \epsilon)}
\end{align}
Therefore
\begin{align}
	\delta V_Y(r)&=(Res\ \delta V_Y)(iM)-\frac{e^2}{2\pi^2 r}\int_{-\infty}^{\infty}\dd{q}\frac{q\ Im[\Pi(q^2+i\epsilon)]}{(q^2-M^2)^2}e^{-qr}
\end{align}
The residue computed over the path $\Gamma=\sum \gamma$ is zero. Utilizing this result along with the imaginary part \ref{Im.Pi} the previous expression takes the form
\begin{align}
	\delta V_Y(r)&=-\frac{e^2 g^2}{192\pi^3 r}\int_{m+M}^{\infty}\dd{q}\frac{e^{-q r}}{q(q^2-M^2)^2}\qty[\qty((m-M)^2-q^2)\qty((m+M)^2-q^2)]^{3/2}
\end{align}
Finally, the corrected potential is ($q=t(M+m)$)
\begin{align}\label{resul:potential}
	V(r)=-\frac{e^2}{4\pi}\qty(\frac{e^{-M r}}{r}+
	\frac{g^2(m+M)^2}{3\times2^4\pi^2}\frac{1}{r}\int_{1}^{\infty}\dd{t}
	F\qty(m/M,t)\qty(t^2-1)^{3/2}\frac{e^{-(m+M)rt}}{t}
	)
\end{align}
with
\begin{align}\label{resul:potential.F}
	F\qty(m/M,t)=\qty(t^2-\qty(\frac{M-m}{M+m})^2)^{3/2}\qty(t^2-\qty(\frac{M}{M+m})^2)^{-2}
\end{align}
It is not clear how to compute the $t$ integral in full analytic form, but some doable simplifications can extract analytical information in some limiting cases.

%%----------------------------------------------------------------------------------------
\section{Analysis of the results}
\begin{figure}[h!]
	\centering
	\includegraphics[width=0.8\textwidth]{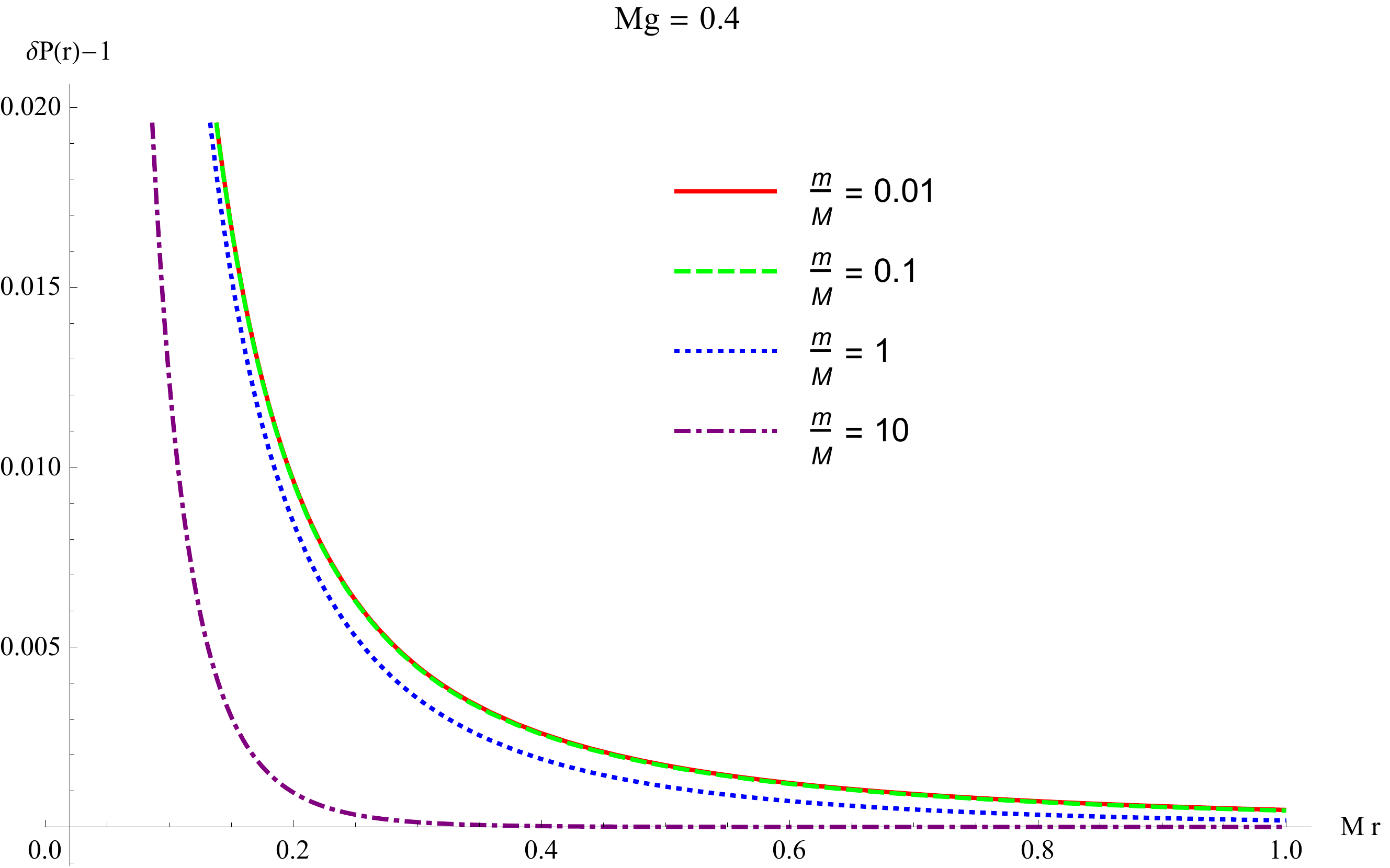}
	\caption{Graph of the exact expression of $\delta P-1$ (equation \ref{resul:delta.P}) for varying values of $m/M$. The used values are $g M=0.4$ and $Mr\in \qty(0.02,1)$.}
	\label{quantum.deviation.with.mass.variation}
\end{figure}

%----------------------------------------------------------------------------------------
Equation \ref{resul:potential} can be rewritten as 
\begin{align}
	V(r)&=-\frac{e^2}{4\pi}\frac{e^{-Mr}}{r}\delta P(Mr, gM, m/M)\label{resul:V.delta.P}\\
	\delta P(Mr, gM, m/M)&=1+\frac{\qty(gM)^2\qty(1+\frac{m}{M})^2}{48\pi^2}\int_{1}^{\infty}\dd t F\qty(m/M,t)\qty(t^2-1)^{3/2}\frac{e^{-M r\qty[t\qty(1+m/M)-1]}}{t}\label{resul:delta.P}
\end{align}
where $\delta P(Mr, gM, m/M)$, which corresponds to deviations from the Yukawa potential introduced by quantum fluctuations of the axion field, is organized in terms of three dimensionless parameters $(Mr, gM,m/M)$. We remark that all the computations so far do not rely on any specific relationship between these three parameters but, since this is an emergent description of the system, these are effective parameters that are related to each other and fixed by the microscopic physics as previously discussed in section II. Yet, for the sake of simplicity, we will continue to treat these parameters as independent for now.
The graphical representation (for a set of self-consistent parameters described in section \ref{Graph_numerical_integration}) of the quantum deviation (namely $\delta P-1$) is given in figure \ref{quantum.deviation.with.mass.variation}. 
To develop a physical picture, it is useful to analyze the result \ref{resul:delta.P} imposing large mass hierarchies (large axion mass $m\gg M$ and large Proca mass $M\gg m$).    

Each approximation will provide an estimated result that, for additional verification, will be compared against the numerical integration.
%%----------------------------------------------------------------------------------------
\subsection{Asymptotic Approximations}
%----------------------------------------------------------------------------------------
\subsubsection{Small Axion mass}
%----------------------------------------------------------------------------------------
Applying a small axion mass approximation ($M\gg m$) at zero-order in the mass ratio $\frac{m}{M}$, the expression equation \ref{resul:delta.P} simplifies to
\begin{align}
\delta P(Mr,gM)= 1
+\frac{g^2 M^2}{48\pi^2}\int_1^{\infty}\dd{t} \qty(t^2-1)\frac{e^{-M r (t-1)}}{t}+\order{\frac{m}{M}}
\end{align}
Evaluating the integral we obtain 
\begin{align}\label{resul:potential.a.b.small.axion}
\delta P(Mr,gM)\approx 1
+\frac{g^2 M^2}{48\pi^2}
\qty(\qty(\frac{1}{M^2 r^2}+\frac{1}{M r})-e^{Mr}\Gamma(0,Mr))
\end{align}
where $\Gamma(0,Mr)$ is the upper incomplete gamma function\footnote{Defined as $\Gamma(a,x)\equiv\int_x^\infty t^{a-1}e^{-t}\dd{t}$. The asymptotic expression of $\Gamma(0,Mr)$ for $Mr\gg1$ is $\propto \frac{e^{-Mr}}{Mr}$, this cancels the possible problem of the positive exponent $e^{Mr}$ in \ref{resul:potential.a.b.small.axion}.}. The asymptotic approximation results in
\begin{align}\label{deltapy}
\delta P(Mr,gM)\approx
\begin{cases}
1+\frac{g^2 M^2}{24 \pi^2}\frac{1}{(Mr)^2}
; \;\; \text{for} \;\; Mr \gg 1\\
1+\frac{g^2 M^2}{48\pi^2}\qty(\frac{1}{(Mr)^2}+\frac{1}{M r}+\log(e^\gamma M r))
; \;\; \text{for} \;\; Mr \ll 1
\end{cases}
\end{align}
Figure \ref{newgraph1} and \ref{newgraph3} compare the results with the numerical integration without approximations.

%----------------------------------------------------------------------------------------
\subsubsection{Small Proca mass}
\begin{figure}[H]
	\centering
	\begin{subfigure}[b]{0.45\linewidth}
		\includegraphics[width=\linewidth]{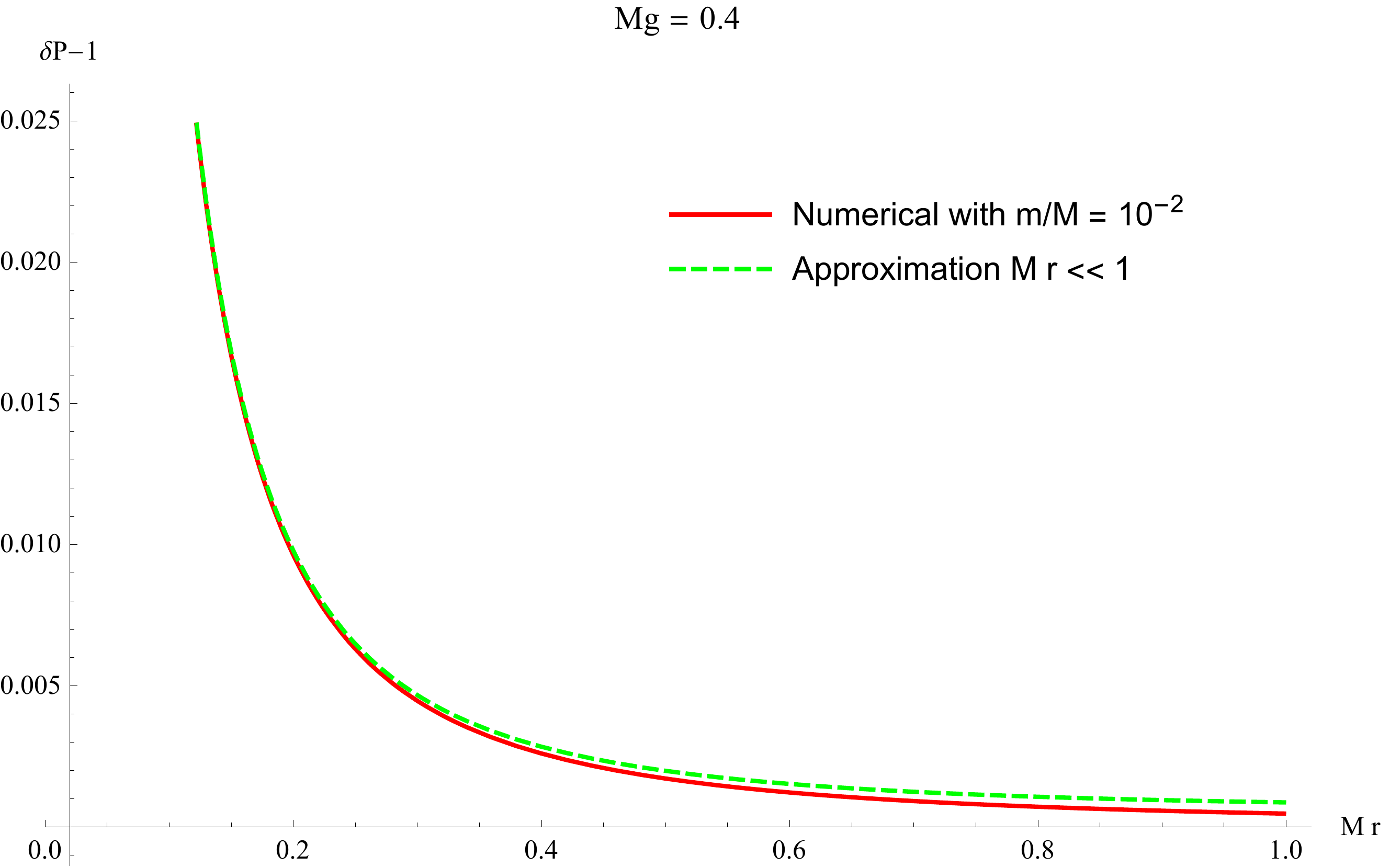}
		\caption{Approximation valid up to $Mr\sim 0.2$.\label{newgraph1}}
	\end{subfigure}
	\begin{subfigure}[b]{0.45\linewidth}
		\includegraphics[width=\linewidth]{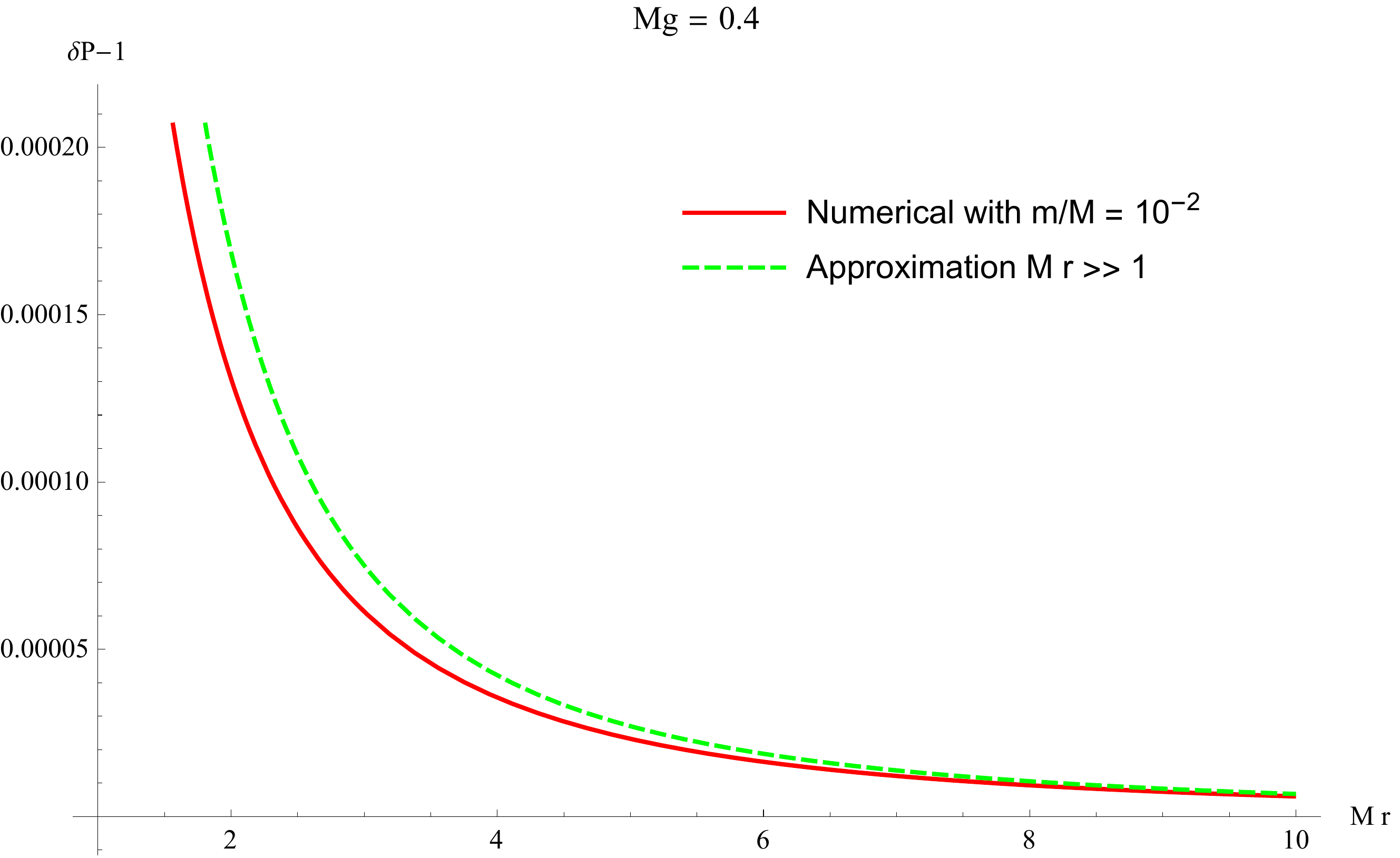}
		\caption{Approximation valid starting at $Mr\sim 8$.\label{newgraph3}}
	\end{subfigure}
	\begin{subfigure}[b]{0.45\linewidth}
		\includegraphics[width=\linewidth]{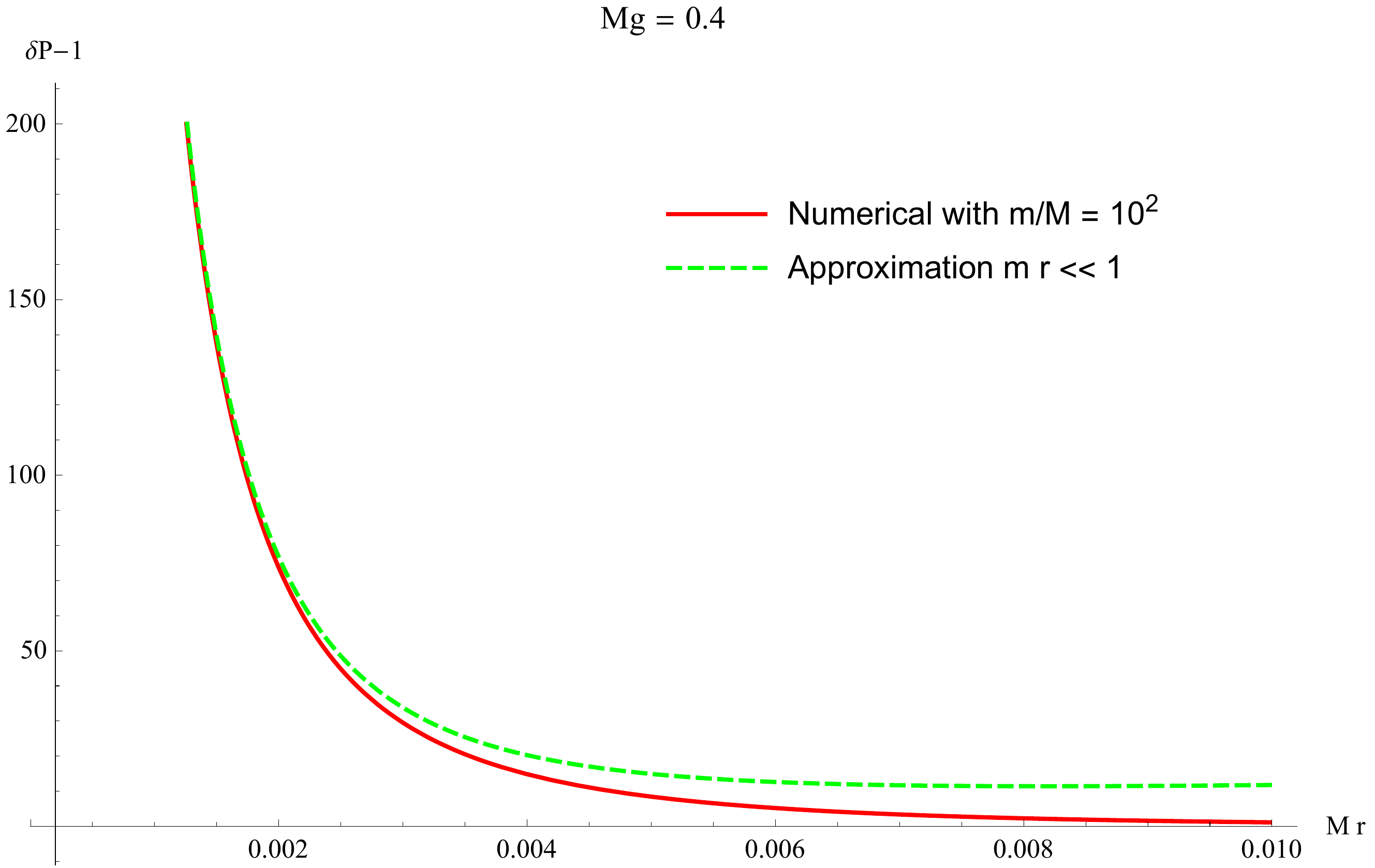}
		\caption{Approximation valid up to $Mr\sim 0.002\to mr\sim 0.2$. \label{newgraph5}}
	\end{subfigure}
	\begin{subfigure}[b]{0.45\linewidth}
		\includegraphics[width=\linewidth]{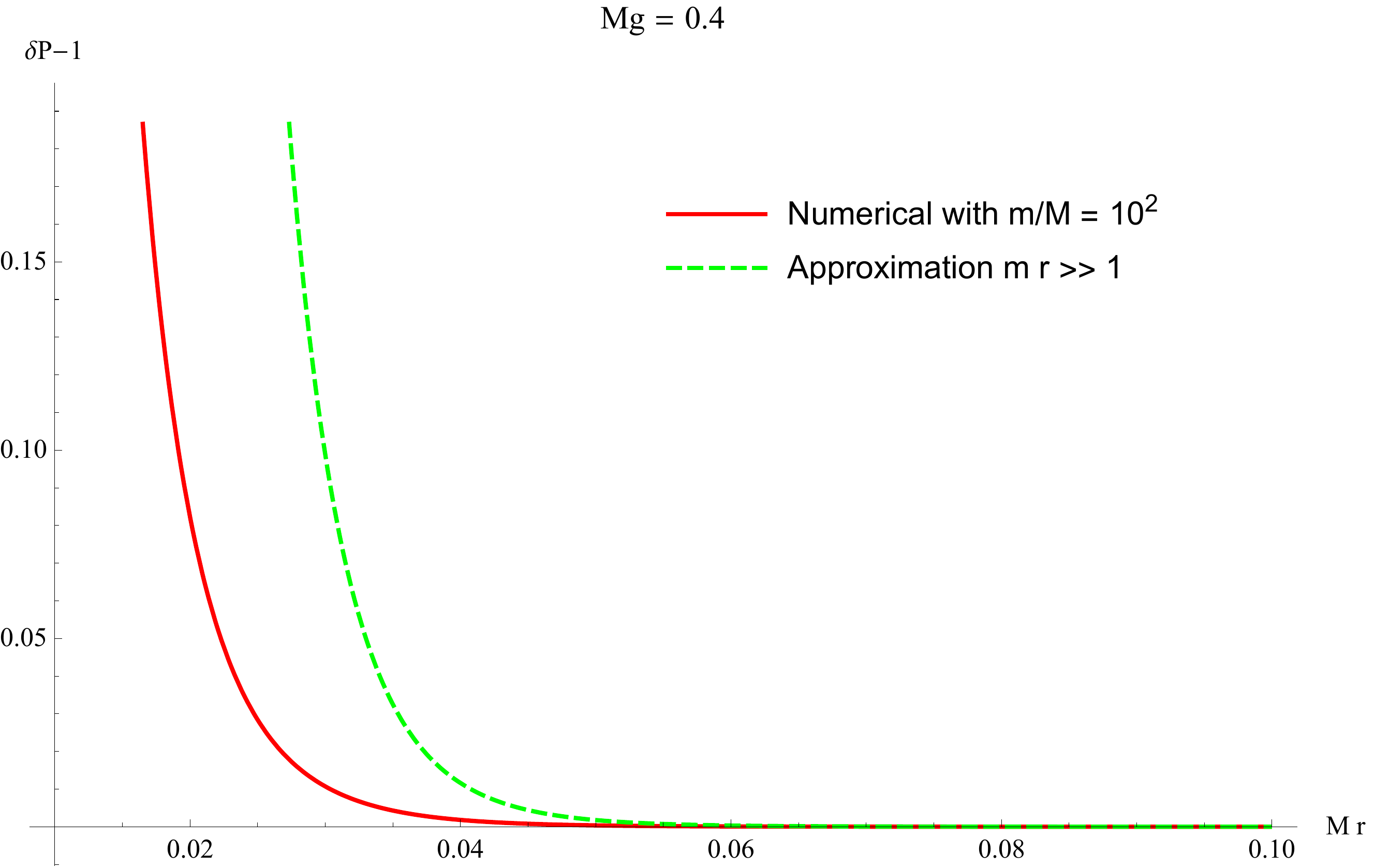}
		\caption{Approximation valid starting at $Mr\sim 0.06\to mr\sim 6$. \label{newgraph7}}
	\end{subfigure}
	\caption{Plot of $\delta P(r)-1$ (the deviation from the standard value) as a function of $M r$ with $g M=0.4$. The red line is the numerical integration plot of \ref{resul:delta.P}. Respectively; \ref{newgraph1} and \ref{newgraph3} represent the approximated function \ref{deltapy} with $M r\ll 1$ and with $M r\gg 1$. Moreover, \ref{newgraph5} and \ref{newgraph7} represent \ref{deltapy2} with $m r\ll 1$ and with $m r\gg 1$ (the relation $\frac{m}{M}=10$ is also used to express all functions in terms of $Mr$). The estimate of the region of validity of the approximations is read directly from the graph.}\label{GraphMlargerthenm1}
\end{figure}

%----------------------------------------------------------------------------------------
In the case of a small Proca mass, in comparison with the axion mass ($M\ll m$), eq. \ref{resul:delta.P} gives
\begin{align}\label{smallProcamass.integral}
\delta P(Mr,gM,m/M)&=1+\frac{g^2m^2}{48\pi^2}
\int_1^\infty \dd{t} \left(
\frac{\qty(t^2-1)^3}{t^5}
+2\frac{M}{m}\frac{\qty(t^2-1)^2\qty(t^2+2)}{t^5}
\nonumber\right.\\	
&\left.\qquad
+\frac{M^2}{m^2}\frac{\qty(t^2-1)\qty{2+3t^2+t^6}}{t^7}	
\right)e^{-mrt-Mr(t-1)}+\order{\dfrac{M}{m}}^3
\end{align}
This integral, that can be computed analytically, but does not bring any valuable insight, is expressed in the appendix  \ref{fullformafterintegration}. Employing the asymptotic expansion in these expressions results in\footnote{Note that every term in this expression can be expressed in terms of $(Mr, gM, m/ M)$.}
\begin{align}\label{deltapy2}
\delta P(Mr,gM,m/M)\approx
\begin{cases}
1+\frac{g^2m^2}{\pi^2}
\qty(
\frac{1}{\qty(m+M)^4 r^2}+\frac{M}{m}\frac{1}{\qty(m+M)^3 r}+\frac{M^2}{m^2}\frac{1}{4\qty(m+M)^2}
)\frac{e^{-mr}}{r^2}	
; \;\; \text{for} \;\; mr \gg 1\\
1+\frac{g^2m^2}{48\pi^2}\left(
\frac{1}{\qty(mr)^2}+\frac{3}{4}
+\frac{M}{m}\qty(\frac{1}{mr}-3)
+3\log\qty(e^{\gamma}\qty(m+M)r)
\right.\\
\qquad\left.		
+\frac{M^2}{m^2}\qty(\frac{11}{12}+\log\qty(e^{\gamma}\qty( m+M)r))
\right)
; \;\; \text{for} \;\; mr \ll 1
\end{cases}
\end{align}
The graphs \ref{newgraph5} and \ref{newgraph7} represents the comparison between the full numerical integration and the approximations. Note that this result is consistent with the massless photon limit that was examined in \cite{Villalba-Chavez:2018eql}.
%---------
% Figure 2
%---------

%----------------------------------------------------------------------------------------
\subsection{Mass relations and London penetration length}
\begin{figure}[h!]
	\centering
	\includegraphics[width=0.8\textwidth]{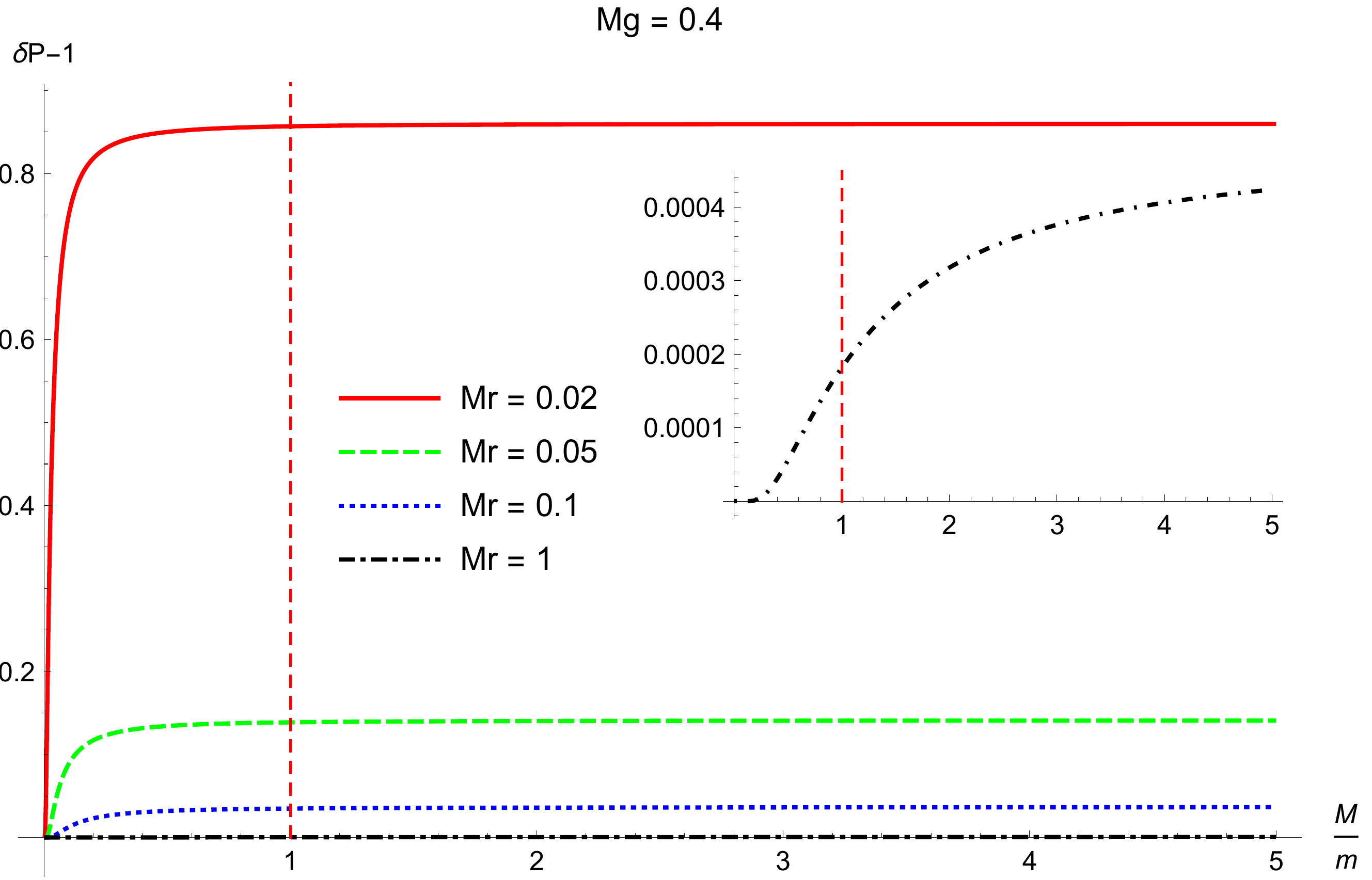}
	\caption{Graph of the exact expression of $\delta P-1$ \eqref{resul:delta.P} as a function of $M/m$ for varying values of the distance scale with fixed $Mg=0.4$. The inserted graph is the zoom of the curve $Mr=1$. The red vertical red line $M/m=1$ separates the region with $M/m<1$ and $M/m>1$.}
	\label{graph.mass.variation}
\end{figure}

\begin{figure}[h!]
	\centering
	\includegraphics[width=0.8\textwidth]{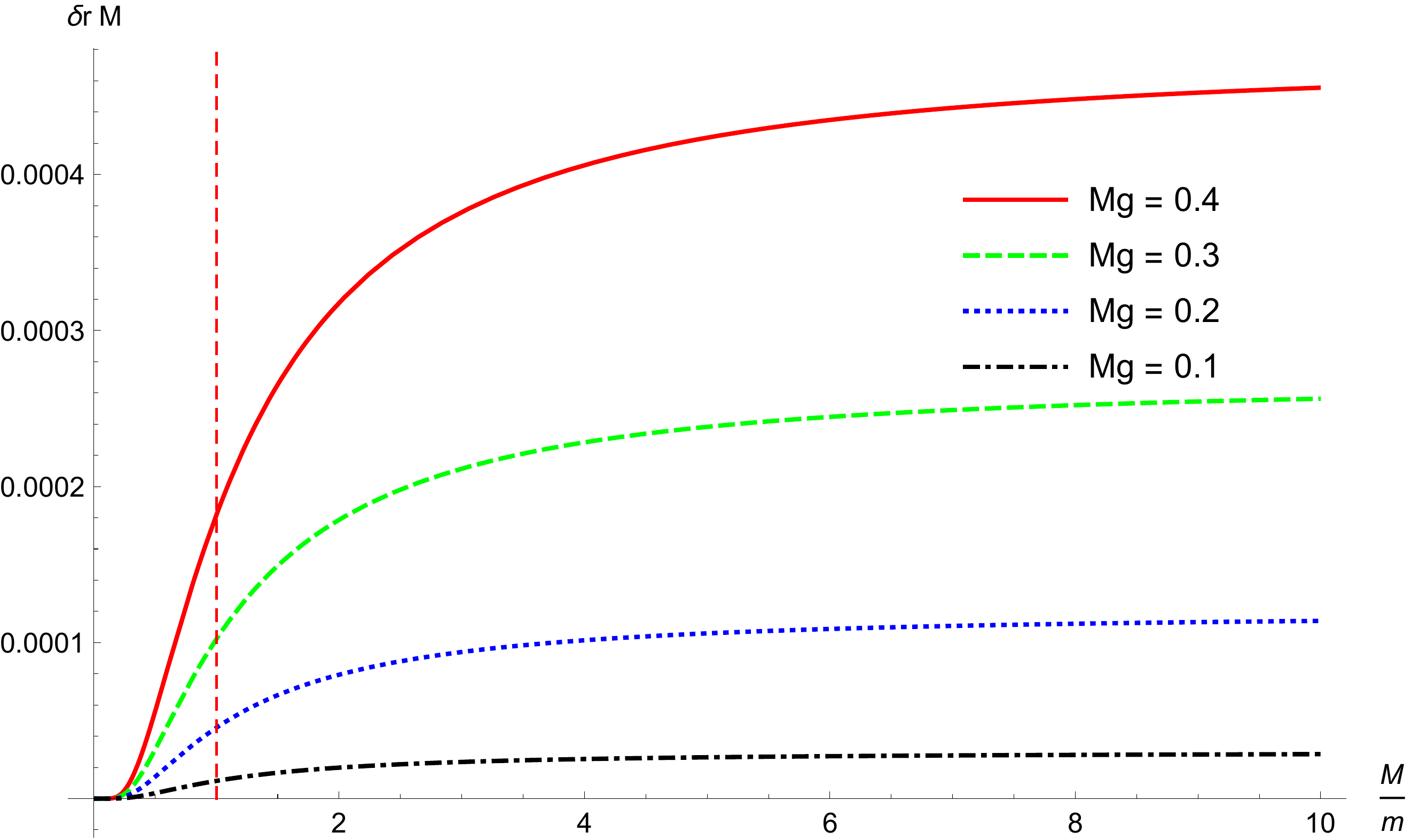}
	\caption{Plot of the expression $M\delta r$ \eqref{resul:delta.P} as a function of $M/m$ with different values of $Mg$. The red vertical red line $M/m=1$ separates the region with $M/m<1$ and $M/m>1$.}
	\label{graph.London.mass.variation}
\end{figure}

Considering the results depicted in figure \ref{graph.mass.variation} (the variation of the quantum correction as the mass ratio changes) we see that as $m$ becomes larger than $M$ quantum corrections becomes less and less important. One can also note that for large distances the corrections are very feeble for any values of the masses. This means that we expect noticeable deviations from the usual London results, due to axion effects, at small penetration distances and large photon mass ($M/m > 1$).  
In fact, we can explore in more details  the variation in the London screening generated by quantum fluctuations of the axion background. To do so, it is useful to redefine \ref{resul:V.delta.P} with an effective mass by
\begin{align}\label{londonmanipulate}
	V(r)=-\frac{e^2}{4\pi}\frac{e^{-r M^{\text{eff}}(Mr,Mg,m/M)}}{r}
\end{align}
so that
\begin{align}\label{lambdaeff}
	M^{\text{eff}}(Mr,Mg,m/M)
	=M-\frac{\log\delta P}{r}
	=M-\frac{\delta P-1}{r}+\order{g^4}
\end{align}
where $\delta P=\delta P(Mr,Mg,m/M)$ is given by \ref{resul:delta.P} and the expansion $\log \left( 1 + a x  \right) \approx a x $ was used. The Yukawa tree level interaction, {\it i.e } $ V_Y (r)= -\frac{e^2}{4\pi} \dfrac{e^{ - M r } }{ r } $, defines the London length $\lambda_L$ as the damping coefficient of the exponential via $e^{-M\lambda_L} = e^{-1}$, or equivalently, $\lambda_{ L } = \dfrac{ 1 }{ M }$. We can expect that this term receives quantum corrections that can be writtten in the form
\begin{align}\label{M.eff.r.eff}
	r^{eff}M^{eff}(r^{eff})=1+\order{g^2}
\end{align}
that is a transcendental equation, but it is possible to solve by considering that 
\begin{align}\label{r.eff}
	r^{eff}=\lambda_L+\delta r+\order{g^4}
\end{align}
where $\delta r\in\order{g^2}$ and $M\lambda_L=1$ resulting in\footnote{This expression was obtained by expanding $e^{-M r^{eff}\qty[t\qty(1+m/M)-1]}$ (with the use of equation \ref{r.eff}) and keeping terms of $\order{g^0}$ since the whole integral is of $\order{g^2}$. Note that this follows the same spirit of the renormalization of the charge in QED.}
\begin{align}
	M\delta r=\frac{(gM)^2\qty(1+\frac{m}{M})^2}{48\pi^2}e^1\int_1^\infty \dd{t} F(m/M,t)\qty(t^2-1)^{3/2}\frac{e^{-t\qty(1+\frac{m}{M})}}{t}+\order{g^4}
\end{align}
This is the term $\order{g^2}$ (leading contribution) expected in \ref{M.eff.r.eff} and is independent of the scale $Mr$. We can see in graph \ref{graph.London.mass.variation} the shift $\delta r$ (in units of $M$) in the London penetration length as a function of the mass ratio $\frac{M}{m}$. As stated before, the axionic effects are more relevant  for large photon mass.

%	Conclusions
%----------------------------------------------------------------------------------------
%----------------------------------------------------------------------------------------
\section{Conclusions}
%----------------------------------------------------------------------------------------
In this work, we investigated the axion-electromagnetic theory obtained from the electromagnetic response of a Dirac semimetal with a quartic pairing instability. The pairing effectively induces the dynamical formation of a charged chiral condensate whose phases fluctuations give rise to an effective axionic excitation along with a longitudinal mode for the photon excitations through the Higgs mechanism. As mentioned, the Axion mass is related to charge density waves of the fermionic condensate, and the resulting fully gapped system describes an axionic superconductor.

We also investigated the two-point function of the massive photon excitation considering one-loop axionic corrections and found that these corrections naturally induce a modification of typical electromagnetic interaction at short distances. Consequently, in the asymptotic limit, the effective theory is Yukawa-type (Proca) representing an usual superconductor.

To be more precise, based on the discussion of section VI-B, these modifications should play a role for average lengths below $r\sim1.25\ nm$ in systems with characteristic electromagnetic interaction length of $M \sim \left( 50\ nm \right)^{ - 1 } $ \cite{Kittel2004}. We remark however that this is an educated guess based on average experimental values to illustrate the range of parameters that would give a physically significant effect.

The maximum possible value for the correction occurs when the axion mass is lesser or equal to the photon mass. Oppositely, as the Axion mass becomes larger, i.e. the field becomes harder to excite, the quantum fluctuations become closer to the non-perturbed value ($M^{eff}\sim M$). This reasoning is based, partially, on the fact that axion emission, by a decay process of $\gamma\to\gamma \theta$, is not possible.

As stated before, in the course of our calculations we regarded the effective parameters $m$, $M$, and $g$ as unrelated quantities. However, if we take into account the microscopic origin, as discussed in section II, we must consider the connection between them and the microscopic parameters $\lambda$ and $v$. The scaling relations are $g\sim \frac{1}{\lambda^2 v^3}$, $M\sim \lambda^2 v^3$ and $m\sim \frac{1}{\lambda}$, that can be reduced to $g\sim \frac{1}{M}$ and $m\sim\sqrt{\frac{v^3}{M}}$. These relations are compatible with the range of values considered in our analysis since the perturbative computations are valid for $gM<1$. Our results also indicate that Axionic effects are more prominent when $M>m$. 

In conclusion, that since the order of magnitude of distance adopted in section VI-B is appropriate to thin-films physics, the electromagnetic screening properties (by the corrected London length) of thin-films constituted by superconducting Dirac materials could be sensible to the described effects in preceding sections. This is a possible probe to the quantum effects due to axionic coupling. However, it is important to stress that, at this stage, the explicit connection between our findings and the aforementioned discussion as well as the practical applicability or even feasibility to real condensed matter systems is lacking, being a topic for further investigation.

\section*{Acknowledgements}
The authors would like to thank the Brazilian agencies CNPq (Conselho Nacional de Desenvolvimento Cient{\'\i}fico e Tecnol\'ogico) and FAPERJ (Funda{\c c}{\~a}o de Amparo à Pesquisa do Estado do Rio de Janeiro) for financial support. This study was financed in part by the CAPES (Coordena{\c c}{\~a}o de Aperfei{\c c}oamento de Pessoal de N{\'\i}vel Superior - Brasil), Finance Code 001. M.S.G. is a level 2 CNPq researcher under Contract No. 307801/2017-9.

\appendix
%----------------------------------------------------------------------------------------
\section{}
%----------------------------------------------------------------------------------------
\label{bare.to.renor.ghost.elimination}
\subsection{Renormalization}
The ``bare" field and parameters must be replaced by the renormalized ones, that is, we must replace the following quantities in equations (\ref{initial.action}-\ref{g2pertub}).
\begin{align}
	\begin{matrix}
		\overline{a}_\mu\to A_\mu && f_{\mu\nu} \to F_{\mu\nu} && \overline{\theta} \to \theta\\
		\overline{M}\to M && \overline m \to m && \overline{g}\to g\\	
		\overline m_1 \to m_1 && \overline C_{\theta} \to C_{\theta} &&
		\overline m_s \to m_s\\
		\overline M_2 \to M_2 &&
		\overline C_{f} \to C_{f}&&
		\overline m_{gh} \to m_{gh} \\
		\overline C_{4} \to C_{4}&&
		\overline M_2 \to M_2&&
		\overline C_{a \theta} \to C_{a \theta}&&
		\overline m^2_{\theta f} \to m^2_{\theta f}
	\end{matrix}
\end{align}
The renormalized action becomes
\begin{align}
	S_R=\int d^4 x
	\left(
	\mathcal{L}_{\text{Proca}}+\mathcal{L}_{\text{axion}}
	+\mathcal{L}_{\text{interaction}}
	+ \mathcal{L}_{\theta g^2}+ \mathcal{L}_{a g^2}+ \mathcal{L}_{a\theta g^2}
	\right)
\end{align}
with the Proca and axion Lagrangians being
\begin{subequations}\label{lagran.free.append}
	\begin{align}
		\mathcal{L}_{\text{Proca}}&=-\frac{1}{4}Z_3 F_{\mu\nu}F^{\mu\nu}+\frac{1}{2}Z_M M^{2}A_{\mu}A^{\mu}\\
		\mathcal{L}_{\text{axion}}&=\frac{1}{2}Z_\theta \partial_\mu \theta\partial^\mu \theta-\frac{1}{2}Z_m m^{2}\theta^{2}
	\end{align}
\end{subequations}
, the interaction term ($\order{g^1}$)
\begin{align}\label{lagran.int.append}
	\mathcal{L}_{\text{interaction}}&=\frac{1}{4}Z_g g\theta\tilde{F}_{\mu\nu}F^{\mu\nu}
\end{align}
and the next-to-leading ($\order{g^2}$)
\begin{align}
	\mathcal{L}_{\theta g^2}&=-\frac{1}{2}Z_{m2} \theta^2 m_1^2+\frac{1}{2}Z_{\theta2}  C_\theta(\partial  \theta)^2+\frac{Z_{s}}{2 m_s^2}(\partial_\mu  \theta)\square(\partial^\mu  \theta)\\
	\mathcal{L}_{a g^2}&=\frac{1}{2}Z_{M2} M_1^2  A^2-\frac{1}{4}Z_{f} C_{f}F^2+\frac{Z_{gh}}{2  m_{gh}^2}(\partial F)^2+\frac{1}{4!}Z_{4}C_{4} A^4  -\frac{1}{4!}Z_{5}\frac{ A^2}{ M_{2}^2} F^2\\
	\mathcal{L}_{a\theta g^2}&=-\frac{1}{4}Z_{a \theta} C_{a\theta}  \theta^2  A^2 +\frac{1}{4}Z_{\theta f}\frac{ \theta^2}{ m_{\theta f}^2} F^2
\end{align}
Some terms in the last equation can be incorporated in the free section plus a modification $\order{g^4}$ and can be ignored since this is outside the scope of our $1$-loop computation. The redefinition is
\begin{align}
	\begin{matrix}
		Z_3\to\qty(1-C_{f})Z_3, && M^2 Z_M\to\qty(M^2-M_1^2)Z_M \\	
		Z_\theta\to\qty(1-C_\theta)Z_\theta, && m^2 Z_m\to\qty(m^2-m_1^2)Z_m	
	\end{matrix}
\end{align}
This changes the $g^2$ part to
\begin{subequations}\label{lagran.order.g2}
	\begin{align}
		\mathcal{L}_{\theta g^2}&=\frac{Z_{s}}{2 m_s^2}(\partial_\mu  \theta)\square(\partial^\mu  \theta)\\
		\mathcal{L}_{a g^2}&=\frac{Z_{gh}}{2  m_{gh}^2}(\partial F)^2+\frac{1}{4!}Z_{4}C_{4} A^4  -\frac{1}{4!}Z_{5}\frac{ A^2}{ M_{2}^2} F^2\\
		\mathcal{L}_{a\theta g^2}&=-\frac{1}{4}Z_{a \theta} C_{a\theta}  \theta^2  A^2 +\frac{1}{4}Z_{\theta f}\frac{ \theta^2}{ m_{\theta f}^2} F^2
	\end{align}
\end{subequations}
%----------------------------------------------------------------------------------------
\subsection{Parameters relation}
%----------------------------------------------------------------------------------------
Now we can derive the connection between the ``bare" parameters and the renormalized ones. Using the kinetic prescription, $A^\mu=Z_3^{-1/2}a^\mu$ and $\theta=Z_\theta^{-1/2} \overline{\theta}$, results in the following relations
\begin{align}
	\label{def:bare.to.renor}
	\begin{matrix}
		\overline{M}=M\left(\frac{Z_M}{Z_3}\right)^{1/2}, && \overline{m}=m\left(\frac{Z_m}{Z_\theta}\right)^{1/2}, && \overline{m}_{gh}=m_{gh}\left(\frac{Z_3}{Z_{gh}}\right)^{1/2},\\
		\overline{m}_s=m_s\left(\frac{Z_\theta}{Z_s}\right)^{1/2}, && \overline{C}_{4}=C_{4}\frac{Z_3}{Z_{a^4}^{1/2}}, && \overline{M}_2=M_2\frac{Z_3}{Z_5^{1/2}},\\
		\overline{m}_{\theta f}=m_{\theta f}\left(\frac{Z_3 Z_\theta}{Z_{\theta f}}\right)^{1/2}, && \overline{C}_{a\theta}=C_{a\theta}\left(\frac{Z_{a\theta}}{Z_3 Z_{\theta}}\right)^{1/2}, &&\overline{g}=g\frac{Z_g}{Z_3^{1/2}Z_\theta^{1/2}}.
	\end{matrix}
\end{align}
%----------------------------------------------------------------------------------------
\subsection{Ghost elimination process}
%----------------------------------------------------------------------------------------
The action composed of \ref{lagran.free.append} and \ref{lagran.order.g2} still exhibits the problem of higher derivative contributions to the free sector. These contributions can not be an oversight because they will modify the free propagator by introducing a new ``mass pole'' for the pseudoscalar  and massive vector field causing the introduction of non-physical states. These terms can not be absorbed in a parameter shift because they carry a $\square^2$ (or in momentum space, $~p^4$) dependency. It is possible to eliminate these terms using a field redefinition
\begin{align}
	\theta \to \theta-\frac{\square}{2m_{s}^2}\theta\qquad
	A_\mu \to A_\mu-\frac{\square}{2m_{gh}^2}A_\mu
\end{align}
, any extra term will be of order $g^4$ and can be ignored as it is outside the wanted perturbative accuracy. This process is described in the appendix of \citep{Villalba-Chavez:2018eql} (and reference within it). The final product is the original Lagrangian minus the ghosts generating terms but with the reward of retaining their counter-term. This is crucial to the renormalization process in section IV.B. The resulting action is
\begin{align}
	\mathcal{L}_R&=\mathcal{L}_{\text{Proca}}+\mathcal{L}_{\text{axion}}+\mathcal{L}_{\text{interaction}}+\frac{\delta_{s}}{2 m_s^2}(\partial_\mu  \theta)\square(\partial^\mu  \theta)+\frac{\delta_{gh}}{2  m_{gh}^2}(\partial F)^2
	+\mathcal{L}_{4\gamma}+\mathcal{L}_{2\gamma,2\theta}
\end{align}
with
\begin{subequations}
	\begin{align}
		\mathcal{L}_{4\gamma}&=\frac{1}{4!}Z_{4}C_{4} A^4-\frac{1}{4!}Z_5\frac{ A^2}{ M_2^2} F^2\\
		\mathcal{L}_{2\gamma,2\theta}&=-\frac{1}{4}Z_{a \theta} C_{a\theta}  \theta^2  A^2 +\frac{1}{4}Z_{\theta f}\frac{ \theta^2}{ m_{\theta f}^2} F^2
	\end{align}
\end{subequations}
along with equations \ref{lagran.free.append} and \ref{lagran.int.append}.  
%%----------------------------------------------------------------------------------------
%	Some mathematical general details
%----------------------------------------------------------------------------------------
\section{Mathematical details}
%----------------------------------------------------------------------------------------
\subsection{Graph numerical integration}
\label{Graph_numerical_integration}
%----------------------------
In order to analyze how the effective theory changes as the parameters are modified is convenient to introduce a set of dimensionless combinations.  The dimensional parameters $\qty(m,M,g,r)$ can be arranged in in three dimensionless terms: $M r$ (distance scale), $\frac{m}{M}$ (mass ratio scale), and $g M$ (coupling scale). Notice that in this parametrization a larger (smaller) axion mass, than Proca mass, translates to $m/M>1$ $(0<m/M<1)$.

This results in the polarization \ref{resul:delta.P} taking the form $\delta P\qty(Mr,g M,m/M)=1+f\qty(Mr,g M,m/M)$ with
\begin{align}\label{condition1}
	f\qty(Mr,gM,m/M):=\frac{\qty(gM)^2\qty(1+\frac{m}{M})^2}{48\pi^2}\int_{1}^{\infty}\dd{t} F\qty(m/M,t)\qty(t^2-1)^{3/2}\frac{e^{-M r\qty[t\qty(1+m/M)-1]}}{t}
\end{align}
Any specification of $\qty(Mr,g M,m/M)$ must be consistent with the perturbation theory and physical experimental ranges. To be compatible with perturbation theory they must obey
\begin{align}\label{delta.P.inequality}
	f\qty(Mr,gM,m/M)<1
\end{align}
This inequality can be studied graphically using numerical inputs of phenomenological characteristic scales. 

The outline of the analysis is; It is possible to define a $f\qty(M_0 r_0,(g M)_{crit},m_0/M_0)$ with some $M g _{\text{crit}}$. In order to keep the perturbative analysis consistent in a given range $ M r \in [ (Mr)_{min} , (Mr)_{max} ] $ and $m/M \in [ 0 , (m/M)_{ max }  ]$, it is sufficient to choose a value $ M g < (M g)_{\text{crit}} $ that can be determined either numerically or graphically using the values of $(M r, m/M) = ((M r)_{min}, 0)$.

\begin{figure}[h!]
	\centering
	\begin{subfigure}[b]{0.45\linewidth}
		\includegraphics[width=0.8\textwidth]{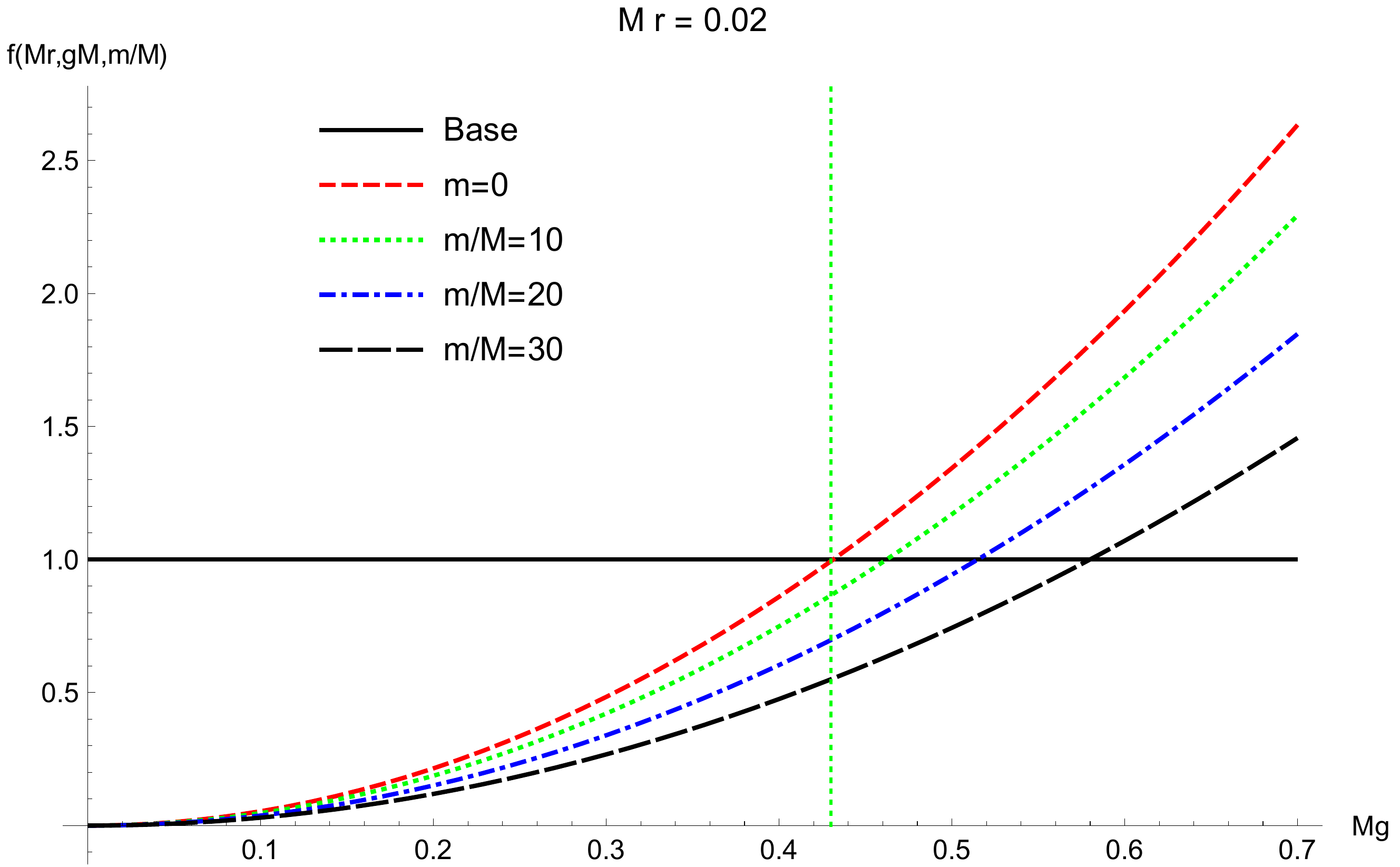}
		\caption{The numerical plot of left and right hand sides of \eqref{delta.P.inequality}, for $Mr = 0.02$. Note that, the critical values of $gM$ that keeps the perturbative analysis valid increases with $\frac{m}{M}$.}
		\label{setaG}
	\end{subfigure}
	\begin{subfigure}[b]{0.45\linewidth}
		\includegraphics[width=0.8\textwidth]{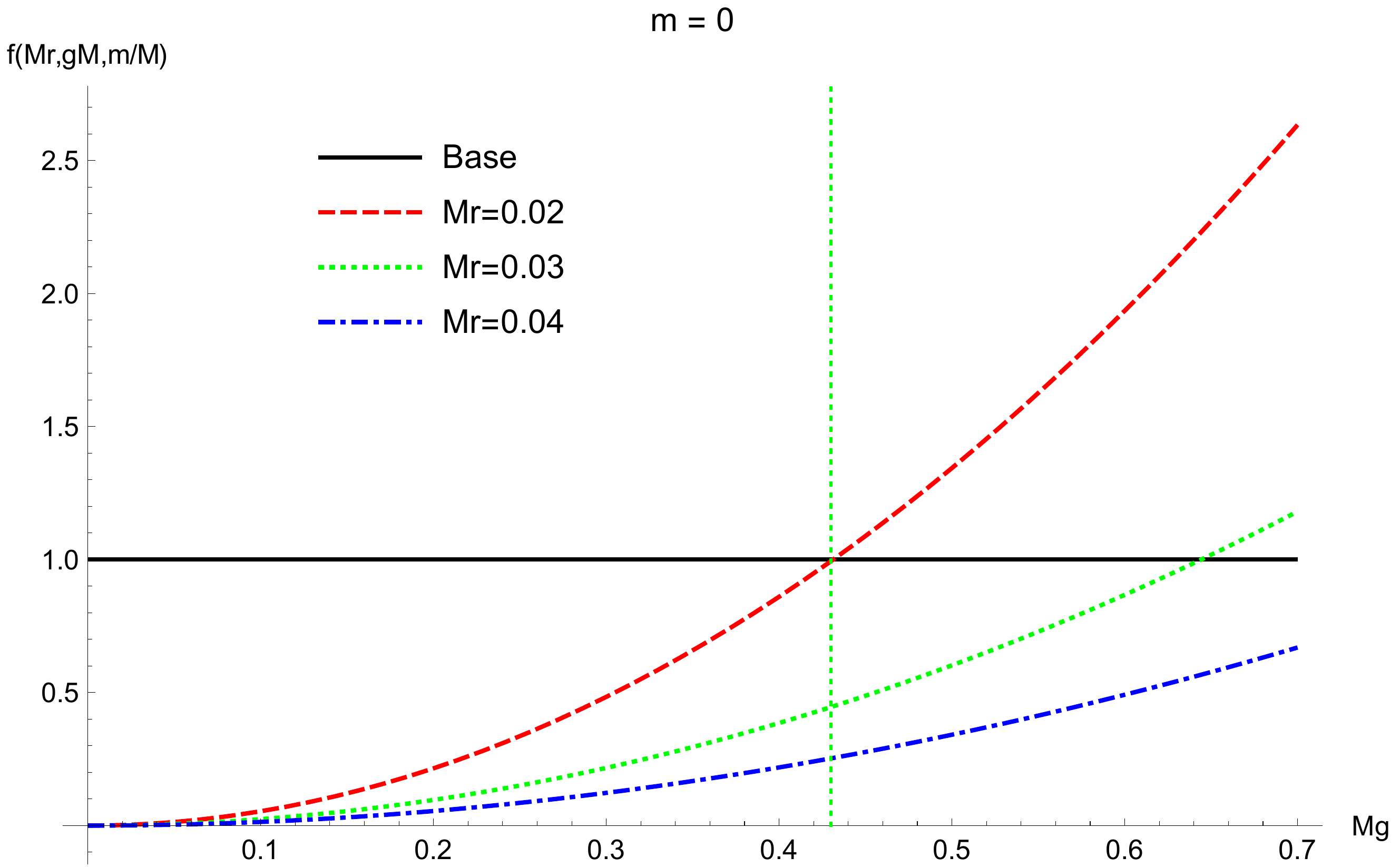}
		\caption{The numerical plot of left and right hand sides of \eqref{delta.P.inequality}, for $m = 0$. Note that the critical values of $gM$ also increases considerably as one makes slightly modifications on $Mr$.}
		\label{setaG2}
	\end{subfigure}
	\caption{Numerical analysis of the inequality \eqref{delta.P.inequality}. The black line in 1 represents the upper bound and the vertical dashed line is the critical value $Mg=0.43$.}
\end{figure}

Considering a separation in the order of nanometers and take the London length usually found in superconductors (that ranges from $\lambda_{ L } \sim 50 \, nm $ to $\sim 500 \, nm $ \cite{Kittel2004}) as a representative scale for the photon's mass. Theoretically, this setup is experimental feasible since it consists of a thin film of superconductor. Now consider length scales running from $ r \sim 1 \,  nm $ to $ r \sim 50 \,  nm $. This choice of $M\sim 1/50\ nm^{-1}$ lead to $M r \in [0.02,1]$.  In order to get a consistent value of $ M g $ for {\it any } $ M r $ greater than the lower bound it is sufficient to solve \eqref{condition1} for $\qty(Mr,m/M)=\qty(0.02,0) $. Graphically it can be read from figure \ref{setaG} ) that this is true for $\eval{Mg}_{crit}\approx 0.43$. This sets the typical length scale above which the perturbative analysis breaks and our model is not reliable anymore. 

%-------------------------------
% Figures
%-------------------------------
% Figure 1
%---------

%----------------------------------------------------------------------------------------
\subsection{Full expression}
\label{fullformafterintegration}
%----------------------------------------------------------------------------------------
The full integral of \ref{smallProcamass.integral} is
\begin{align}\label{smallProcamass.integrated}
	\delta P (r)=1+\frac{g^2}{\pi^2}\qty[
	e^{-mr}Fun1
	+e^{Mr} Fun2
	]+\order{\frac{M}{m}^1}
\end{align}
with
\begin{align}
	Fun1&=\frac{r^3 (m+M)^3 \left(15 m^2-60 m M+17 M^2\right)}{17280}-\frac{r^2 (m+M)^2 \left(5 m^2-20 m M+7 M^2\right)}{5760}
	\nonumber\\
	&\qquad	
	-\frac{r (m+M) \left(85 m^2-160 m M+81 M^2\right)}{2880}+\frac{1}{576} \left(15 m^2-24 m M+11 M^2\right)
	\nonumber\\
	&\qquad	
	+\frac{M^2 r^5 (m+M)^5}{17280}-\frac{M^2 r^4 (m+M)^4}{17280}+\frac{m+M}{48 r}+\frac{1}{48 r^2}\label{smallProcamass.integrated.2}\\
	Fun2&=\frac{r^4 (m+M)^4 \left(m^2-4 m M+M^2\right) \text{Ei}(-(m+M) r)}{1152}-\frac{1}{32} r^2 \left(m^2-M^2\right)^2 \text{Ei}(-(m+M) r)
	\nonumber\\
	&\qquad	
	+\frac{1}{48} \left(3 m^2+M^2\right) \text{Ei}(-(m+M) r)+\frac{M^2 r^6 (m+M)^6 \text{Ei}(-(m+M) r)}{17280}\label{smallProcamass.integrated.3}
\end{align}
%----------------------------------------------------------------------------------------
\newpage
\bibliography{axionprocaM}

\begin{thebibliography}{55}%
\makeatletter
\providecommand \@ifxundefined [1]{%
 \@ifx{#1\undefined}
}%
\providecommand \@ifnum [1]{%
 \ifnum #1\expandafter \@firstoftwo
 \else \expandafter \@secondoftwo
 \fi
}%
\providecommand \@ifx [1]{%
 \ifx #1\expandafter \@firstoftwo
 \else \expandafter \@secondoftwo
 \fi
}%
\providecommand \natexlab [1]{#1}%
\providecommand \enquote  [1]{``#1''}%
\providecommand \bibnamefont  [1]{#1}%
\providecommand \bibfnamefont [1]{#1}%
\providecommand \citenamefont [1]{#1}%
\providecommand \href@noop [0]{\@secondoftwo}%
\providecommand \href [0]{\begingroup \@sanitize@url \@href}%
\providecommand \@href[1]{\@@startlink{#1}\@@href}%
\providecommand \@@href[1]{\endgroup#1\@@endlink}%
\providecommand \@sanitize@url [0]{\catcode `\\12\catcode `\$12\catcode
  `\&12\catcode `\#12\catcode `\^12\catcode `\_12\catcode `\%12\relax}%
\providecommand \@@startlink[1]{}%
\providecommand \@@endlink[0]{}%
\providecommand \url  [0]{\begingroup\@sanitize@url \@url }%
\providecommand \@url [1]{\endgroup\@href {#1}{\urlprefix }}%
\providecommand \urlprefix  [0]{URL }%
\providecommand \Eprint [0]{\href }%
\providecommand \doibase [0]{https://doi.org/}%
\providecommand \selectlanguage [0]{\@gobble}%
\providecommand \bibinfo  [0]{\@secondoftwo}%
\providecommand \bibfield  [0]{\@secondoftwo}%
\providecommand \translation [1]{[#1]}%
\providecommand \BibitemOpen [0]{}%
\providecommand \bibitemStop [0]{}%
\providecommand \bibitemNoStop [0]{.\EOS\space}%
\providecommand \EOS [0]{\spacefactor3000\relax}%
\providecommand \BibitemShut  [1]{\csname bibitem#1\endcsname}%
\let\auto@bib@innerbib\@empty
%</preamble>
\bibitem [{\citenamefont {Weinberg}(1975)}]{Weinberg:1975ui}%
  \BibitemOpen
  \bibfield  {author} {\bibinfo {author} {\bibfnamefont {S.}~\bibnamefont
  {Weinberg}},\ }\href {https://doi.org/10.1103/PhysRevD.11.3583} {\bibfield
  {journal} {\bibinfo  {journal} {Phys. Rev. D}\ }\textbf {\bibinfo {volume}
  {11}},\ \bibinfo {pages} {3583} (\bibinfo {year} {1975})}\BibitemShut
  {NoStop}%
\bibitem [{\citenamefont {'t~Hooft}(1976)}]{tHooft:1976snw}%
  \BibitemOpen
  \bibfield  {author} {\bibinfo {author} {\bibfnamefont {G.}~\bibnamefont
  {'t~Hooft}},\ }\href {https://doi.org/10.1103/PhysRevD.14.3432} {\bibfield
  {journal} {\bibinfo  {journal} {Phys. Rev. D}\ }\textbf {\bibinfo {volume}
  {14}},\ \bibinfo {pages} {3432} (\bibinfo {year} {1976})},\ \bibinfo {note}
  {[Erratum: Phys.Rev.D 18, 2199 (1978)]}\BibitemShut {NoStop}%
\bibitem [{\citenamefont {'t~Hooft}\ \emph {et~al.}(1980)\citenamefont
  {'t~Hooft}, \citenamefont {Itzykson}, \citenamefont {Jaffe}, \citenamefont
  {Lehmann}, \citenamefont {Mitter}, \citenamefont {Singer},\ and\
  \citenamefont {Stora}}]{tHooft:1980xss}%
  \BibitemOpen
  \bibinfo {editor} {\bibfnamefont {G.}~\bibnamefont {'t~Hooft}}, \bibinfo
  {editor} {\bibfnamefont {C.}~\bibnamefont {Itzykson}}, \bibinfo {editor}
  {\bibfnamefont {A.}~\bibnamefont {Jaffe}}, \bibinfo {editor} {\bibfnamefont
  {H.}~\bibnamefont {Lehmann}}, \bibinfo {editor} {\bibfnamefont
  {P.}~\bibnamefont {Mitter}}, \bibinfo {editor} {\bibfnamefont
  {I.}~\bibnamefont {Singer}},\ and\ \bibinfo {editor} {\bibfnamefont
  {R.}~\bibnamefont {Stora}},\ eds.,\ \href
  {https://doi.org/10.1007/978-1-4684-7571-5} {\emph {\bibinfo {title} {{Recent
  Developments in Gauge Theories. Proceedings, Nato Advanced Study Institute,
  Cargese, France, August 26 - September 8, 1979}}}},\ Vol.~\bibinfo {volume}
  {59}\ (\bibinfo {year} {1980})\BibitemShut {NoStop}%
\bibitem [{\citenamefont {'t~Hooft}(1986)}]{tHooft:1986ooh}%
  \BibitemOpen
  \bibfield  {author} {\bibinfo {author} {\bibfnamefont {G.}~\bibnamefont
  {'t~Hooft}},\ }\href {https://doi.org/10.1016/0370-1573(86)90117-1}
  {\bibfield  {journal} {\bibinfo  {journal} {Phys. Rept.}\ }\textbf {\bibinfo
  {volume} {142}},\ \bibinfo {pages} {357} (\bibinfo {year}
  {1986})}\BibitemShut {NoStop}%
\bibitem [{\citenamefont {Peccei}\ and\ \citenamefont
  {Quinn}(1977{\natexlab{a}})}]{Peccei:1977hh}%
  \BibitemOpen
  \bibfield  {author} {\bibinfo {author} {\bibfnamefont {R.}~\bibnamefont
  {Peccei}}\ and\ \bibinfo {author} {\bibfnamefont {H.~R.}\ \bibnamefont
  {Quinn}},\ }\href {https://doi.org/10.1103/PhysRevLett.38.1440} {\bibfield
  {journal} {\bibinfo  {journal} {Phys. Rev. Lett.}\ }\textbf {\bibinfo
  {volume} {38}},\ \bibinfo {pages} {1440} (\bibinfo {year}
  {1977}{\natexlab{a}})}\BibitemShut {NoStop}%
\bibitem [{\citenamefont {Peccei}\ and\ \citenamefont
  {Quinn}(1977{\natexlab{b}})}]{Peccei:1977ur}%
  \BibitemOpen
  \bibfield  {author} {\bibinfo {author} {\bibfnamefont {R.}~\bibnamefont
  {Peccei}}\ and\ \bibinfo {author} {\bibfnamefont {H.~R.}\ \bibnamefont
  {Quinn}},\ }\href {https://doi.org/10.1103/PhysRevD.16.1791} {\bibfield
  {journal} {\bibinfo  {journal} {Phys. Rev. D}\ }\textbf {\bibinfo {volume}
  {16}},\ \bibinfo {pages} {1791} (\bibinfo {year}
  {1977}{\natexlab{b}})}\BibitemShut {NoStop}%
\bibitem [{\citenamefont {Peccei}(2008)}]{Peccei:2006as}%
  \BibitemOpen
  \bibfield  {author} {\bibinfo {author} {\bibfnamefont {R.}~\bibnamefont
  {Peccei}},\ }\href {https://doi.org/10.1007/978-3-540-73518-2\_1} {\bibfield
  {journal} {\bibinfo  {journal} {Lect. Notes Phys.}\ }\textbf {\bibinfo
  {volume} {741}},\ \bibinfo {pages} {3} (\bibinfo {year} {2008})},\ \Eprint
  {https://arxiv.org/abs/hep-ph/0607268} {arXiv:hep-ph/0607268} \BibitemShut
  {NoStop}%
\bibitem [{\citenamefont {Weinberg}(1978)}]{Weinberg:1977ma}%
  \BibitemOpen
  \bibfield  {author} {\bibinfo {author} {\bibfnamefont {S.}~\bibnamefont
  {Weinberg}},\ }\href {https://doi.org/10.1103/PhysRevLett.40.223} {\bibfield
  {journal} {\bibinfo  {journal} {Phys. Rev. Lett.}\ }\textbf {\bibinfo
  {volume} {40}},\ \bibinfo {pages} {223} (\bibinfo {year} {1978})}\BibitemShut
  {NoStop}%
\bibitem [{\citenamefont {Wilczek}(1978)}]{Wilczek:1977pj}%
  \BibitemOpen
  \bibfield  {author} {\bibinfo {author} {\bibfnamefont {F.}~\bibnamefont
  {Wilczek}},\ }\href {https://doi.org/10.1103/PhysRevLett.40.279} {\bibfield
  {journal} {\bibinfo  {journal} {Phys. Rev. Lett.}\ }\textbf {\bibinfo
  {volume} {40}},\ \bibinfo {pages} {279} (\bibinfo {year} {1978})}\BibitemShut
  {NoStop}%
\bibitem [{\citenamefont {Kim}\ and\ \citenamefont
  {Carosi}(2010)}]{Kim:2008hd}%
  \BibitemOpen
  \bibfield  {author} {\bibinfo {author} {\bibfnamefont {J.~E.}\ \bibnamefont
  {Kim}}\ and\ \bibinfo {author} {\bibfnamefont {G.}~\bibnamefont {Carosi}},\
  }\href {https://doi.org/10.1103/RevModPhys.82.557} {\bibfield  {journal}
  {\bibinfo  {journal} {Rev. Mod. Phys.}\ }\textbf {\bibinfo {volume} {82}},\
  \bibinfo {pages} {557} (\bibinfo {year} {2010})},\ \bibinfo {note} {[Erratum:
  Rev.Mod.Phys. 91, 049902 (2019)]},\ \Eprint {https://arxiv.org/abs/0807.3125}
  {arXiv:0807.3125 [hep-ph]} \BibitemShut {NoStop}%
\bibitem [{\citenamefont {Braaten}\ and\ \citenamefont
  {Zhang}(2019)}]{Braaten:2019knj}%
  \BibitemOpen
  \bibfield  {author} {\bibinfo {author} {\bibfnamefont {E.}~\bibnamefont
  {Braaten}}\ and\ \bibinfo {author} {\bibfnamefont {H.}~\bibnamefont
  {Zhang}},\ }\href {https://doi.org/10.1103/RevModPhys.91.041002} {\bibfield
  {journal} {\bibinfo  {journal} {Rev. Mod. Phys.}\ }\textbf {\bibinfo {volume}
  {91}},\ \bibinfo {pages} {041002} (\bibinfo {year} {2019})}\BibitemShut
  {NoStop}%
\bibitem [{\citenamefont {Kim}(1979)}]{Kim:1979if}%
  \BibitemOpen
  \bibfield  {author} {\bibinfo {author} {\bibfnamefont {J.~E.}\ \bibnamefont
  {Kim}},\ }\href {https://doi.org/10.1103/PhysRevLett.43.103} {\bibfield
  {journal} {\bibinfo  {journal} {Phys. Rev. Lett.}\ }\textbf {\bibinfo
  {volume} {43}},\ \bibinfo {pages} {103} (\bibinfo {year} {1979})}\BibitemShut
  {NoStop}%
\bibitem [{\citenamefont {Shifman}\ \emph {et~al.}(1980)\citenamefont
  {Shifman}, \citenamefont {Vainshtein},\ and\ \citenamefont
  {Zakharov}}]{Shifman:1979if}%
  \BibitemOpen
  \bibfield  {author} {\bibinfo {author} {\bibfnamefont {M.~A.}\ \bibnamefont
  {Shifman}}, \bibinfo {author} {\bibfnamefont {A.}~\bibnamefont
  {Vainshtein}},\ and\ \bibinfo {author} {\bibfnamefont {V.~I.}\ \bibnamefont
  {Zakharov}},\ }\href {https://doi.org/10.1016/0550-3213(80)90209-6}
  {\bibfield  {journal} {\bibinfo  {journal} {Nucl. Phys. B}\ }\textbf
  {\bibinfo {volume} {166}},\ \bibinfo {pages} {493} (\bibinfo {year}
  {1980})}\BibitemShut {NoStop}%
\bibitem [{\citenamefont {Dine}\ \emph {et~al.}(1981)\citenamefont {Dine},
  \citenamefont {Fischler},\ and\ \citenamefont {Srednicki}}]{Dine:1981rt}%
  \BibitemOpen
  \bibfield  {author} {\bibinfo {author} {\bibfnamefont {M.}~\bibnamefont
  {Dine}}, \bibinfo {author} {\bibfnamefont {W.}~\bibnamefont {Fischler}},\
  and\ \bibinfo {author} {\bibfnamefont {M.}~\bibnamefont {Srednicki}},\ }\href
  {https://doi.org/10.1016/0370-2693(81)90590-6} {\bibfield  {journal}
  {\bibinfo  {journal} {Phys. Lett. B}\ }\textbf {\bibinfo {volume} {104}},\
  \bibinfo {pages} {199} (\bibinfo {year} {1981})}\BibitemShut {NoStop}%
\bibitem [{\citenamefont {Marsh}(2016)}]{Marsh:2015xka}%
  \BibitemOpen
  \bibfield  {author} {\bibinfo {author} {\bibfnamefont {D.~J.~E.}\
  \bibnamefont {Marsh}},\ }\href
  {https://doi.org/10.1016/j.physrep.2016.06.005} {\bibfield  {journal}
  {\bibinfo  {journal} {Phys. Rept.}\ }\textbf {\bibinfo {volume} {643}},\
  \bibinfo {pages} {1} (\bibinfo {year} {2016})},\ \Eprint
  {https://arxiv.org/abs/1510.07633} {arXiv:1510.07633 [astro-ph.CO]}
  \BibitemShut {NoStop}%
\bibitem [{\citenamefont {Wilczek}(1987)}]{Wilczek:1987mv}%
  \BibitemOpen
  \bibfield  {author} {\bibinfo {author} {\bibfnamefont {F.}~\bibnamefont
  {Wilczek}},\ }\href {https://doi.org/10.1103/PhysRevLett.58.1799} {\bibfield
  {journal} {\bibinfo  {journal} {Phys. Rev. Lett.}\ }\textbf {\bibinfo
  {volume} {58}},\ \bibinfo {pages} {1799} (\bibinfo {year}
  {1987})}\BibitemShut {NoStop}%
\bibitem [{\citenamefont {Moore}(2010)}]{2010Natur.464..194M}%
  \BibitemOpen
  \bibfield  {author} {\bibinfo {author} {\bibfnamefont {J.}~\bibnamefont
  {Moore}},\ }\href {https://doi.org/10.1038/nature08916} {\bibfield  {journal}
  {\bibinfo  {journal} {Nature}\ }\textbf {\bibinfo {volume} {464}},\ \bibinfo
  {pages} {194} (\bibinfo {year} {2010})}\BibitemShut {NoStop}%
\bibitem [{\citenamefont {Hasan}\ and\ \citenamefont
  {Kane}(2010)}]{Hasan:2010xy}%
  \BibitemOpen
  \bibfield  {author} {\bibinfo {author} {\bibfnamefont {M.}~\bibnamefont
  {Hasan}}\ and\ \bibinfo {author} {\bibfnamefont {C.}~\bibnamefont {Kane}},\
  }\href {https://doi.org/10.1103/RevModPhys.82.3045} {\bibfield  {journal}
  {\bibinfo  {journal} {Rev. Mod. Phys.}\ }\textbf {\bibinfo {volume} {82}},\
  \bibinfo {pages} {3045} (\bibinfo {year} {2010})},\ \Eprint
  {https://arxiv.org/abs/1002.3895} {arXiv:1002.3895 [cond-mat.mes-hall]}
  \BibitemShut {NoStop}%
\bibitem [{\citenamefont {Qi}\ and\ \citenamefont {Zhang}(2011)}]{Qi:2011zya}%
  \BibitemOpen
  \bibfield  {author} {\bibinfo {author} {\bibfnamefont {X.~L.}\ \bibnamefont
  {Qi}}\ and\ \bibinfo {author} {\bibfnamefont {S.~C.}\ \bibnamefont {Zhang}},\
  }\href {https://doi.org/10.1103/RevModPhys.83.1057} {\bibfield  {journal}
  {\bibinfo  {journal} {Rev. Mod. Phys.}\ }\textbf {\bibinfo {volume} {83}},\
  \bibinfo {pages} {1057} (\bibinfo {year} {2011})},\ \Eprint
  {https://arxiv.org/abs/1008.2026} {arXiv:1008.2026 [cond-mat.mes-hall]}
  \BibitemShut {NoStop}%
\bibitem [{\citenamefont {Hasan}\ and\ \citenamefont
  {Moore}(2011)}]{Hasan2011ThreeDimensionalTI}%
  \BibitemOpen
  \bibfield  {author} {\bibinfo {author} {\bibfnamefont {M.~Z.}\ \bibnamefont
  {Hasan}}\ and\ \bibinfo {author} {\bibfnamefont {J.~E.}\ \bibnamefont
  {Moore}},\ }\href {https://doi.org/10.1146/annurev-conmatphys-062910-140432}
  {\bibfield  {journal} {\bibinfo  {journal} {Annual Review of Condensed Matter
  Physics}\ }\textbf {\bibinfo {volume} {2}},\ \bibinfo {pages} {55} (\bibinfo
  {year} {2011})},\ \Eprint
  {https://arxiv.org/abs/https://doi.org/10.1146/annurev-conmatphys-062910-140432}
  {https://doi.org/10.1146/annurev-conmatphys-062910-140432} \BibitemShut
  {NoStop}%
\bibitem [{\citenamefont {Qi}\ \emph {et~al.}(2008)\citenamefont {Qi},
  \citenamefont {Hughes},\ and\ \citenamefont {Zhang}}]{Qi:2008ew}%
  \BibitemOpen
  \bibfield  {author} {\bibinfo {author} {\bibfnamefont {X.-L.}\ \bibnamefont
  {Qi}}, \bibinfo {author} {\bibfnamefont {T.}~\bibnamefont {Hughes}},\ and\
  \bibinfo {author} {\bibfnamefont {S.-C.}\ \bibnamefont {Zhang}},\ }\href
  {https://doi.org/10.1103/PhysRevB.78.195424} {\bibfield  {journal} {\bibinfo
  {journal} {Phys. Rev. B}\ }\textbf {\bibinfo {volume} {78}},\ \bibinfo
  {pages} {195424} (\bibinfo {year} {2008})},\ \Eprint
  {https://arxiv.org/abs/0802.3537} {arXiv:0802.3537 [cond-mat.mes-hall]}
  \BibitemShut {NoStop}%
\bibitem [{\citenamefont {{von Klitzing}}\ \emph {et~al.}(2020)\citenamefont
  {{von Klitzing}}, \citenamefont {Chakraborty}, \citenamefont {Kim},
  \citenamefont {Madhavan}, \citenamefont {Dai}, \citenamefont {McIver},
  \citenamefont {Tokura}, \citenamefont {Savary}, \citenamefont {Smirnova},
  \citenamefont {Rey}, \citenamefont {Felser}, \citenamefont {Gooth},\ and\
  \citenamefont {Qi}}]{quantum.hall.effect.40.years}%
  \BibitemOpen
  \bibfield  {author} {\bibinfo {author} {\bibfnamefont {K.}~\bibnamefont {{von
  Klitzing}}}, \bibinfo {author} {\bibfnamefont {T.}~\bibnamefont
  {Chakraborty}}, \bibinfo {author} {\bibfnamefont {P.}~\bibnamefont {Kim}},
  \bibinfo {author} {\bibfnamefont {V.}~\bibnamefont {Madhavan}}, \bibinfo
  {author} {\bibfnamefont {X.}~\bibnamefont {Dai}}, \bibinfo {author}
  {\bibfnamefont {J.}~\bibnamefont {McIver}}, \bibinfo {author} {\bibfnamefont
  {Y.}~\bibnamefont {Tokura}}, \bibinfo {author} {\bibfnamefont
  {L.}~\bibnamefont {Savary}}, \bibinfo {author} {\bibfnamefont
  {D.}~\bibnamefont {Smirnova}}, \bibinfo {author} {\bibfnamefont
  {A.}~\bibnamefont {Rey}}, \bibinfo {author} {\bibfnamefont {C.}~\bibnamefont
  {Felser}}, \bibinfo {author} {\bibfnamefont {J.}~\bibnamefont {Gooth}},\ and\
  \bibinfo {author} {\bibfnamefont {X.}~\bibnamefont {Qi}},\ }\href
  {https://doi.org/10.1038/s42254-020-0209-1} {\bibfield  {journal} {\bibinfo
  {journal} {Nature Reviews Physics}\ }\textbf {\bibinfo {volume} {2}},\
  \bibinfo {pages} {397} (\bibinfo {year} {2020})}\BibitemShut {NoStop}%
\bibitem [{\citenamefont {Fukushima}\ \emph {et~al.}(2008)\citenamefont
  {Fukushima}, \citenamefont {Kharzeev},\ and\ \citenamefont
  {Warringa}}]{Fukushima_2008}%
  \BibitemOpen
  \bibfield  {author} {\bibinfo {author} {\bibfnamefont {K.}~\bibnamefont
  {Fukushima}}, \bibinfo {author} {\bibfnamefont {D.~E.}\ \bibnamefont
  {Kharzeev}},\ and\ \bibinfo {author} {\bibfnamefont {H.~J.}\ \bibnamefont
  {Warringa}},\ }\bibfield  {journal} {\bibinfo  {journal} {Physical Review D}\
  }\textbf {\bibinfo {volume} {78}},\ \href
  {https://doi.org/10.1103/physrevd.78.074033} {10.1103/physrevd.78.074033}
  (\bibinfo {year} {2008})\BibitemShut {NoStop}%
\bibitem [{\citenamefont {{Armitage}}\ \emph {et~al.}(2018)\citenamefont
  {{Armitage}}, \citenamefont {{Mele}},\ and\ \citenamefont
  {{Vishwanath}}}]{2018RvMP...90a5001A}%
  \BibitemOpen
  \bibfield  {author} {\bibinfo {author} {\bibfnamefont {N.~P.}\ \bibnamefont
  {{Armitage}}}, \bibinfo {author} {\bibfnamefont {E.~J.}\ \bibnamefont
  {{Mele}}},\ and\ \bibinfo {author} {\bibfnamefont {A.}~\bibnamefont
  {{Vishwanath}}},\ }\href {https://doi.org/10.1103/RevModPhys.90.015001}
  {\bibfield  {journal} {\bibinfo  {journal} {Reviews of Modern Physics}\
  }\textbf {\bibinfo {volume} {90}},\ \bibinfo {eid} {015001} (\bibinfo {year}
  {2018})},\ \Eprint {https://arxiv.org/abs/1705.01111} {arXiv:1705.01111
  [cond-mat.str-el]} \BibitemShut {NoStop}%
\bibitem [{\citenamefont {{Yan}}\ and\ \citenamefont
  {{Felser}}(2017)}]{2017ARCMP...8..337Y}%
  \BibitemOpen
  \bibfield  {author} {\bibinfo {author} {\bibfnamefont {B.}~\bibnamefont
  {{Yan}}}\ and\ \bibinfo {author} {\bibfnamefont {C.}~\bibnamefont
  {{Felser}}},\ }\href
  {https://doi.org/10.1146/annurev-conmatphys-031016-025458} {\bibfield
  {journal} {\bibinfo  {journal} {Annual Review of Condensed Matter Physics}\
  }\textbf {\bibinfo {volume} {8}},\ \bibinfo {pages} {337} (\bibinfo {year}
  {2017})},\ \Eprint {https://arxiv.org/abs/1611.04182} {arXiv:1611.04182
  [cond-mat.mtrl-sci]} \BibitemShut {NoStop}%
\bibitem [{\citenamefont {Nielsen}\ and\ \citenamefont
  {Ninomiya}(1981)}]{Nielsen:1981hk}%
  \BibitemOpen
  \bibfield  {author} {\bibinfo {author} {\bibfnamefont {H.~B.}\ \bibnamefont
  {Nielsen}}\ and\ \bibinfo {author} {\bibfnamefont {M.}~\bibnamefont
  {Ninomiya}},\ }\href {https://doi.org/10.1016/0370-2693(81)91026-1}
  {\bibfield  {journal} {\bibinfo  {journal} {Phys. Lett. B}\ }\textbf
  {\bibinfo {volume} {105}},\ \bibinfo {pages} {219} (\bibinfo {year}
  {1981})}\BibitemShut {NoStop}%
\bibitem [{\citenamefont {Friedan}(1982)}]{Friedan:1982nk}%
  \BibitemOpen
  \bibfield  {author} {\bibinfo {author} {\bibfnamefont {D.}~\bibnamefont
  {Friedan}},\ }\href {https://doi.org/10.1007/BF01403500} {\bibfield
  {journal} {\bibinfo  {journal} {Commun. Math. Phys.}\ }\textbf {\bibinfo
  {volume} {85}},\ \bibinfo {pages} {481} (\bibinfo {year} {1982})}\BibitemShut
  {NoStop}%
\bibitem [{\citenamefont {Kim}\ \emph {et~al.}(2013)\citenamefont {Kim},
  \citenamefont {Kim}, \citenamefont {Wang}, \citenamefont {Sasaki},
  \citenamefont {Satoh}, \citenamefont {Ohnishi}, \citenamefont {Kitaura},
  \citenamefont {Yang},\ and\ \citenamefont {Li}}]{PhysRevLett.111.246603}%
  \BibitemOpen
  \bibfield  {author} {\bibinfo {author} {\bibfnamefont {H.-J.}\ \bibnamefont
  {Kim}}, \bibinfo {author} {\bibfnamefont {K.-S.}\ \bibnamefont {Kim}},
  \bibinfo {author} {\bibfnamefont {J.-F.}\ \bibnamefont {Wang}}, \bibinfo
  {author} {\bibfnamefont {M.}~\bibnamefont {Sasaki}}, \bibinfo {author}
  {\bibfnamefont {N.}~\bibnamefont {Satoh}}, \bibinfo {author} {\bibfnamefont
  {A.}~\bibnamefont {Ohnishi}}, \bibinfo {author} {\bibfnamefont
  {M.}~\bibnamefont {Kitaura}}, \bibinfo {author} {\bibfnamefont
  {M.}~\bibnamefont {Yang}},\ and\ \bibinfo {author} {\bibfnamefont
  {L.}~\bibnamefont {Li}},\ }\href
  {https://doi.org/10.1103/PhysRevLett.111.246603} {\bibfield  {journal}
  {\bibinfo  {journal} {Phys. Rev. Lett.}\ }\textbf {\bibinfo {volume} {111}},\
  \bibinfo {pages} {246603} (\bibinfo {year} {2013})}\BibitemShut {NoStop}%
\bibitem [{\citenamefont {{Zyuzin}}\ and\ \citenamefont
  {{Burkov}}(2012)}]{2012PhRvB..86k5133Z}%
  \BibitemOpen
  \bibfield  {author} {\bibinfo {author} {\bibfnamefont {A.~A.}\ \bibnamefont
  {{Zyuzin}}}\ and\ \bibinfo {author} {\bibfnamefont {A.~A.}\ \bibnamefont
  {{Burkov}}},\ }\href {https://doi.org/10.1103/PhysRevB.86.115133} {\bibfield
  {journal} {\bibinfo  {journal} {\prb}\ }\textbf {\bibinfo {volume} {86}},\
  \bibinfo {eid} {115133} (\bibinfo {year} {2012})},\ \Eprint
  {https://arxiv.org/abs/1206.1868} {arXiv:1206.1868 [cond-mat.mes-hall]}
  \BibitemShut {NoStop}%
\bibitem [{\citenamefont {Carroll}\ \emph {et~al.}(1990)\citenamefont
  {Carroll}, \citenamefont {Field},\ and\ \citenamefont
  {Jackiw}}]{Carroll:1989vb}%
  \BibitemOpen
  \bibfield  {author} {\bibinfo {author} {\bibfnamefont {S.~M.}\ \bibnamefont
  {Carroll}}, \bibinfo {author} {\bibfnamefont {G.~B.}\ \bibnamefont {Field}},\
  and\ \bibinfo {author} {\bibfnamefont {R.}~\bibnamefont {Jackiw}},\ }\href
  {https://doi.org/10.1103/PhysRevD.41.1231} {\bibfield  {journal} {\bibinfo
  {journal} {Phys. Rev. D}\ }\textbf {\bibinfo {volume} {41}},\ \bibinfo
  {pages} {1231} (\bibinfo {year} {1990})}\BibitemShut {NoStop}%
\bibitem [{\citenamefont {Wang}\ and\ \citenamefont
  {Zhang}(2013)}]{Wang:2012bgb}%
  \BibitemOpen
  \bibfield  {author} {\bibinfo {author} {\bibfnamefont {Z.}~\bibnamefont
  {Wang}}\ and\ \bibinfo {author} {\bibfnamefont {S.-C.}\ \bibnamefont
  {Zhang}},\ }\href {https://doi.org/10.1103/PhysRevB.87.161107} {\bibfield
  {journal} {\bibinfo  {journal} {Phys. Rev. B}\ }\textbf {\bibinfo {volume}
  {87}},\ \bibinfo {pages} {161107} (\bibinfo {year} {2013})},\ \Eprint
  {https://arxiv.org/abs/1207.5234} {arXiv:1207.5234 [cond-mat.str-el]}
  \BibitemShut {NoStop}%
\bibitem [{\citenamefont {{Maciejko}}\ and\ \citenamefont
  {{Nandkishore}}(2014)}]{2014PhRvB..90c5126M}%
  \BibitemOpen
  \bibfield  {author} {\bibinfo {author} {\bibfnamefont {J.}~\bibnamefont
  {{Maciejko}}}\ and\ \bibinfo {author} {\bibfnamefont {R.}~\bibnamefont
  {{Nandkishore}}},\ }\href {https://doi.org/10.1103/PhysRevB.90.035126}
  {\bibfield  {journal} {\bibinfo  {journal} {\prb}\ }\textbf {\bibinfo
  {volume} {90}},\ \bibinfo {eid} {035126} (\bibinfo {year} {2014})},\ \Eprint
  {https://arxiv.org/abs/1311.7133} {arXiv:1311.7133 [cond-mat.str-el]}
  \BibitemShut {NoStop}%
\bibitem [{\citenamefont {You}\ \emph {et~al.}(2016)\citenamefont {You},
  \citenamefont {Cho},\ and\ \citenamefont {Hughes}}]{You:2016wbd}%
  \BibitemOpen
  \bibfield  {author} {\bibinfo {author} {\bibfnamefont {Y.}~\bibnamefont
  {You}}, \bibinfo {author} {\bibfnamefont {G.~Y.}\ \bibnamefont {Cho}},\ and\
  \bibinfo {author} {\bibfnamefont {T.~L.}\ \bibnamefont {Hughes}},\ }\href
  {https://doi.org/10.1103/PhysRevB.94.085102} {\bibfield  {journal} {\bibinfo
  {journal} {Phys. Rev. B}\ }\textbf {\bibinfo {volume} {94}},\ \bibinfo
  {pages} {085102} (\bibinfo {year} {2016})},\ \Eprint
  {https://arxiv.org/abs/1605.02734} {arXiv:1605.02734 [cond-mat.str-el]}
  \BibitemShut {NoStop}%
\bibitem [{\citenamefont {{Li}}\ \emph {et~al.}(2010)\citenamefont {{Li}},
  \citenamefont {{Wang}}, \citenamefont {{Qi}},\ and\ \citenamefont
  {{Zhang}}}]{2010NatPh...6..284L}%
  \BibitemOpen
  \bibfield  {author} {\bibinfo {author} {\bibfnamefont {R.}~\bibnamefont
  {{Li}}}, \bibinfo {author} {\bibfnamefont {J.}~\bibnamefont {{Wang}}},
  \bibinfo {author} {\bibfnamefont {X.-L.}\ \bibnamefont {{Qi}}},\ and\
  \bibinfo {author} {\bibfnamefont {S.-C.}\ \bibnamefont {{Zhang}}},\ }\href
  {https://doi.org/10.1038/nphys1534} {\bibfield  {journal} {\bibinfo
  {journal} {Nature Physics}\ }\textbf {\bibinfo {volume} {6}},\ \bibinfo
  {pages} {284} (\bibinfo {year} {2010})},\ \Eprint
  {https://arxiv.org/abs/0908.1537} {arXiv:0908.1537 [cond-mat.other]}
  \BibitemShut {NoStop}%
\bibitem [{\citenamefont {Bednik}\ \emph {et~al.}(2015)\citenamefont {Bednik},
  \citenamefont {Zyuzin},\ and\ \citenamefont {Burkov}}]{PhysRevB.92.035153}%
  \BibitemOpen
  \bibfield  {author} {\bibinfo {author} {\bibfnamefont {G.}~\bibnamefont
  {Bednik}}, \bibinfo {author} {\bibfnamefont {A.~A.}\ \bibnamefont {Zyuzin}},\
  and\ \bibinfo {author} {\bibfnamefont {A.~A.}\ \bibnamefont {Burkov}},\
  }\href {https://doi.org/10.1103/PhysRevB.92.035153} {\bibfield  {journal}
  {\bibinfo  {journal} {Phys. Rev. B}\ }\textbf {\bibinfo {volume} {92}},\
  \bibinfo {pages} {035153} (\bibinfo {year} {2015})}\BibitemShut {NoStop}%
\bibitem [{\citenamefont {Wang}\ \emph {et~al.}(2020)\citenamefont {Wang},
  \citenamefont {Gioia},\ and\ \citenamefont
  {Burkov}}]{PhysRevLett.124.096603}%
  \BibitemOpen
  \bibfield  {author} {\bibinfo {author} {\bibfnamefont {C.}~\bibnamefont
  {Wang}}, \bibinfo {author} {\bibfnamefont {L.}~\bibnamefont {Gioia}},\ and\
  \bibinfo {author} {\bibfnamefont {A.~A.}\ \bibnamefont {Burkov}},\ }\href
  {https://doi.org/10.1103/PhysRevLett.124.096603} {\bibfield  {journal}
  {\bibinfo  {journal} {Phys. Rev. Lett.}\ }\textbf {\bibinfo {volume} {124}},\
  \bibinfo {pages} {096603} (\bibinfo {year} {2020})}\BibitemShut {NoStop}%
\bibitem [{\citenamefont {Cho}\ \emph {et~al.}(2012)\citenamefont {Cho},
  \citenamefont {Bardarson}, \citenamefont {Lu},\ and\ \citenamefont
  {Moore}}]{PhysRevB.86.214514}%
  \BibitemOpen
  \bibfield  {author} {\bibinfo {author} {\bibfnamefont {G.~Y.}\ \bibnamefont
  {Cho}}, \bibinfo {author} {\bibfnamefont {J.~H.}\ \bibnamefont {Bardarson}},
  \bibinfo {author} {\bibfnamefont {Y.-M.}\ \bibnamefont {Lu}},\ and\ \bibinfo
  {author} {\bibfnamefont {J.~E.}\ \bibnamefont {Moore}},\ }\href
  {https://doi.org/10.1103/PhysRevB.86.214514} {\bibfield  {journal} {\bibinfo
  {journal} {Phys. Rev. B}\ }\textbf {\bibinfo {volume} {86}},\ \bibinfo
  {pages} {214514} (\bibinfo {year} {2012})}\BibitemShut {NoStop}%
\bibitem [{\citenamefont {Fulde}\ and\ \citenamefont
  {Ferrell}(1964)}]{PhysRev.135.A550}%
  \BibitemOpen
  \bibfield  {author} {\bibinfo {author} {\bibfnamefont {P.}~\bibnamefont
  {Fulde}}\ and\ \bibinfo {author} {\bibfnamefont {R.~A.}\ \bibnamefont
  {Ferrell}},\ }\href {https://doi.org/10.1103/PhysRev.135.A550} {\bibfield
  {journal} {\bibinfo  {journal} {Phys. Rev.}\ }\textbf {\bibinfo {volume}
  {135}},\ \bibinfo {pages} {A550} (\bibinfo {year} {1964})}\BibitemShut
  {NoStop}%
\bibitem [{\citenamefont {Wei}\ \emph {et~al.}(2014)\citenamefont {Wei},
  \citenamefont {Chao},\ and\ \citenamefont {Aji}}]{PhysRevB.89.014506}%
  \BibitemOpen
  \bibfield  {author} {\bibinfo {author} {\bibfnamefont {H.}~\bibnamefont
  {Wei}}, \bibinfo {author} {\bibfnamefont {S.-P.}\ \bibnamefont {Chao}},\ and\
  \bibinfo {author} {\bibfnamefont {V.}~\bibnamefont {Aji}},\ }\href
  {https://doi.org/10.1103/PhysRevB.89.014506} {\bibfield  {journal} {\bibinfo
  {journal} {Phys. Rev. B}\ }\textbf {\bibinfo {volume} {89}},\ \bibinfo
  {pages} {014506} (\bibinfo {year} {2014})}\BibitemShut {NoStop}%
\bibitem [{\citenamefont {Li}\ and\ \citenamefont
  {Haldane}(2018)}]{PhysRevLett.120.067003}%
  \BibitemOpen
  \bibfield  {author} {\bibinfo {author} {\bibfnamefont {Y.}~\bibnamefont
  {Li}}\ and\ \bibinfo {author} {\bibfnamefont {F.~D.~M.}\ \bibnamefont
  {Haldane}},\ }\href {https://doi.org/10.1103/PhysRevLett.120.067003}
  {\bibfield  {journal} {\bibinfo  {journal} {Phys. Rev. Lett.}\ }\textbf
  {\bibinfo {volume} {120}},\ \bibinfo {pages} {067003} (\bibinfo {year}
  {2018})}\BibitemShut {NoStop}%
\bibitem [{\citenamefont {Scalapino}(2012)}]{RevModPhys.84.1383}%
  \BibitemOpen
  \bibfield  {author} {\bibinfo {author} {\bibfnamefont {D.~J.}\ \bibnamefont
  {Scalapino}},\ }\href {https://doi.org/10.1103/RevModPhys.84.1383} {\bibfield
   {journal} {\bibinfo  {journal} {Rev. Mod. Phys.}\ }\textbf {\bibinfo
  {volume} {84}},\ \bibinfo {pages} {1383} (\bibinfo {year}
  {2012})}\BibitemShut {NoStop}%
\bibitem [{\citenamefont {Braga}\ \emph {et~al.}(2020)\citenamefont {Braga},
  \citenamefont {Guimaraes},\ and\ \citenamefont {Paganelly}}]{Braga_2020}%
  \BibitemOpen
  \bibfield  {author} {\bibinfo {author} {\bibfnamefont {P.}~\bibnamefont
  {Braga}}, \bibinfo {author} {\bibfnamefont {M.}~\bibnamefont {Guimaraes}},\
  and\ \bibinfo {author} {\bibfnamefont {M.}~\bibnamefont {Paganelly}},\ }\href
  {https://doi.org/10.1016/j.aop.2020.168245} {\bibfield  {journal} {\bibinfo
  {journal} {Annals of Physics}\ }\textbf {\bibinfo {volume} {419}},\ \bibinfo
  {pages} {168245} (\bibinfo {year} {2020})}\BibitemShut {NoStop}%
\bibitem [{\citenamefont {Qi}\ \emph {et~al.}(2013)\citenamefont {Qi},
  \citenamefont {Witten},\ and\ \citenamefont {Zhang}}]{PhysRevB.87.134519}%
  \BibitemOpen
  \bibfield  {author} {\bibinfo {author} {\bibfnamefont {X.-L.}\ \bibnamefont
  {Qi}}, \bibinfo {author} {\bibfnamefont {E.}~\bibnamefont {Witten}},\ and\
  \bibinfo {author} {\bibfnamefont {S.-C.}\ \bibnamefont {Zhang}},\ }\href
  {https://doi.org/10.1103/PhysRevB.87.134519} {\bibfield  {journal} {\bibinfo
  {journal} {Phys. Rev. B}\ }\textbf {\bibinfo {volume} {87}},\ \bibinfo
  {pages} {134519} (\bibinfo {year} {2013})}\BibitemShut {NoStop}%
\bibitem [{\citenamefont {Braga}\ \emph {et~al.}(2016)\citenamefont {Braga},
  \citenamefont {Granado}, \citenamefont {Guimaraes},\ and\ \citenamefont
  {Wotzasek}}]{Braga_2016}%
  \BibitemOpen
  \bibfield  {author} {\bibinfo {author} {\bibfnamefont {P.}~\bibnamefont
  {Braga}}, \bibinfo {author} {\bibfnamefont {D.}~\bibnamefont {Granado}},
  \bibinfo {author} {\bibfnamefont {M.}~\bibnamefont {Guimaraes}},\ and\
  \bibinfo {author} {\bibfnamefont {C.}~\bibnamefont {Wotzasek}},\ }\href
  {https://doi.org/10.1016/j.aop.2016.08.005} {\bibfield  {journal} {\bibinfo
  {journal} {Annals of Physics}\ }\textbf {\bibinfo {volume} {374}},\ \bibinfo
  {pages} {1–15} (\bibinfo {year} {2016})}\BibitemShut {NoStop}%
\bibitem [{\citenamefont {Stone}\ and\ \citenamefont
  {Lopes}(2016)}]{PhysRevB.93.174501}%
  \BibitemOpen
  \bibfield  {author} {\bibinfo {author} {\bibfnamefont {M.}~\bibnamefont
  {Stone}}\ and\ \bibinfo {author} {\bibfnamefont {P.~L. e.~S.}\ \bibnamefont
  {Lopes}},\ }\href {https://doi.org/10.1103/PhysRevB.93.174501} {\bibfield
  {journal} {\bibinfo  {journal} {Phys. Rev. B}\ }\textbf {\bibinfo {volume}
  {93}},\ \bibinfo {pages} {174501} (\bibinfo {year} {2016})}\BibitemShut
  {NoStop}%
\bibitem [{\citenamefont {Hansson}\ \emph {et~al.}(2004)\citenamefont
  {Hansson}, \citenamefont {Oganesyan},\ and\ \citenamefont
  {Sondhi}}]{Hansson_2004}%
  \BibitemOpen
  \bibfield  {author} {\bibinfo {author} {\bibfnamefont {T.}~\bibnamefont
  {Hansson}}, \bibinfo {author} {\bibfnamefont {V.}~\bibnamefont {Oganesyan}},\
  and\ \bibinfo {author} {\bibfnamefont {S.}~\bibnamefont {Sondhi}},\ }\href
  {https://doi.org/10.1016/j.aop.2004.05.006} {\bibfield  {journal} {\bibinfo
  {journal} {Annals of Physics}\ }\textbf {\bibinfo {volume} {313}},\ \bibinfo
  {pages} {497–538} (\bibinfo {year} {2004})}\BibitemShut {NoStop}%
\bibitem [{\citenamefont {Hansson}\ \emph {et~al.}(2015)\citenamefont
  {Hansson}, \citenamefont {Kvorning}, \citenamefont {Nair},\ and\
  \citenamefont {Sreejith}}]{Hansson_2015}%
  \BibitemOpen
  \bibfield  {author} {\bibinfo {author} {\bibfnamefont {T.~H.}\ \bibnamefont
  {Hansson}}, \bibinfo {author} {\bibfnamefont {T.}~\bibnamefont {Kvorning}},
  \bibinfo {author} {\bibfnamefont {V.~P.}\ \bibnamefont {Nair}},\ and\
  \bibinfo {author} {\bibfnamefont {G.~J.}\ \bibnamefont {Sreejith}},\
  }\bibfield  {journal} {\bibinfo  {journal} {Physical Review B}\ }\textbf
  {\bibinfo {volume} {91}},\ \href {https://doi.org/10.1103/physrevb.91.075116}
  {10.1103/physrevb.91.075116} (\bibinfo {year} {2015})\BibitemShut {NoStop}%
\bibitem [{\citenamefont {Gallego~Cadavid}\ and\ \citenamefont
  {Rodriguez}(2019)}]{GallegoCadavid:2019zke}%
  \BibitemOpen
  \bibfield  {author} {\bibinfo {author} {\bibfnamefont {A.}~\bibnamefont
  {Gallego~Cadavid}}\ and\ \bibinfo {author} {\bibfnamefont {Y.}~\bibnamefont
  {Rodriguez}},\ }\href {https://doi.org/10.1016/j.physletb.2019.134958}
  {\bibfield  {journal} {\bibinfo  {journal} {Phys. Lett. B}\ }\textbf
  {\bibinfo {volume} {798}},\ \bibinfo {pages} {134958} (\bibinfo {year}
  {2019})},\ \Eprint {https://arxiv.org/abs/1905.10664} {arXiv:1905.10664
  [hep-th]} \BibitemShut {NoStop}%
\bibitem [{\citenamefont {Villalba-Chávez}\ \emph {et~al.}(2018)\citenamefont
  {Villalba-Chávez}, \citenamefont {Golub},\ and\ \citenamefont
  {Müller}}]{Villalba-Chavez:2018eql}%
  \BibitemOpen
  \bibfield  {author} {\bibinfo {author} {\bibfnamefont {S.}~\bibnamefont
  {Villalba-Chávez}}, \bibinfo {author} {\bibfnamefont {A.}~\bibnamefont
  {Golub}},\ and\ \bibinfo {author} {\bibfnamefont {C.}~\bibnamefont
  {Müller}},\ }\href {https://doi.org/10.1103/PhysRevD.98.115008} {\bibfield
  {journal} {\bibinfo  {journal} {Phys. Rev. D}\ }\textbf {\bibinfo {volume}
  {98}},\ \bibinfo {pages} {115008} (\bibinfo {year} {2018})},\ \Eprint
  {https://arxiv.org/abs/1806.10940} {arXiv:1806.10940 [hep-ph]} \BibitemShut
  {NoStop}%
\bibitem [{\citenamefont {Grinstein}\ \emph {et~al.}(2008)\citenamefont
  {Grinstein}, \citenamefont {O'Connell},\ and\ \citenamefont
  {Wise}}]{Grinstein:2007mp}%
  \BibitemOpen
  \bibfield  {author} {\bibinfo {author} {\bibfnamefont {B.}~\bibnamefont
  {Grinstein}}, \bibinfo {author} {\bibfnamefont {D.}~\bibnamefont
  {O'Connell}},\ and\ \bibinfo {author} {\bibfnamefont {M.~B.}\ \bibnamefont
  {Wise}},\ }\href {https://doi.org/10.1103/PhysRevD.77.025012} {\bibfield
  {journal} {\bibinfo  {journal} {Phys. Rev. D}\ }\textbf {\bibinfo {volume}
  {77}},\ \bibinfo {pages} {025012} (\bibinfo {year} {2008})},\ \Eprint
  {https://arxiv.org/abs/0704.1845} {arXiv:0704.1845 [hep-ph]} \BibitemShut
  {NoStop}%
\bibitem [{\citenamefont {Accioly}\ and\ \citenamefont
  {Dias}(2005)}]{Accioly:2005xf}%
  \BibitemOpen
  \bibfield  {author} {\bibinfo {author} {\bibfnamefont {A.}~\bibnamefont
  {Accioly}}\ and\ \bibinfo {author} {\bibfnamefont {M.}~\bibnamefont {Dias}},\
  }\href {https://doi.org/10.1007/s10773-005-4050-1} {\bibfield  {journal}
  {\bibinfo  {journal} {Int. J. Theor. Phys.}\ }\textbf {\bibinfo {volume}
  {44}},\ \bibinfo {pages} {1123} (\bibinfo {year} {2005})},\ \Eprint
  {https://arxiv.org/abs/hep-th/0511242} {arXiv:hep-th/0511242} \BibitemShut
  {NoStop}%
\bibitem [{\citenamefont {van Hees}(2003)}]{vanHees:2003dk}%
  \BibitemOpen
  \bibfield  {author} {\bibinfo {author} {\bibfnamefont {H.}~\bibnamefont {van
  Hees}},\ }\href@noop {} {\  (\bibinfo {year} {2003})},\ \Eprint
  {https://arxiv.org/abs/hep-th/0305076} {arXiv:hep-th/0305076} \BibitemShut
  {NoStop}%
\bibitem [{\citenamefont {Greiner}\ and\ \citenamefont
  {Reinhardt}(1992)}]{Greiner:1992bv}%
  \BibitemOpen
  \bibfield  {author} {\bibinfo {author} {\bibfnamefont {W.}~\bibnamefont
  {Greiner}}\ and\ \bibinfo {author} {\bibfnamefont {J.}~\bibnamefont
  {Reinhardt}},\ }\href@noop {} {\emph {\bibinfo {title} {{Quantum
  electrodynamics}}}}\ (\bibinfo {year} {1992})\BibitemShut {NoStop}%
\bibitem [{\citenamefont {Schwartz}(2014)}]{Schwartz:2013pla}%
  \BibitemOpen
  \bibfield  {author} {\bibinfo {author} {\bibfnamefont {M.~D.}\ \bibnamefont
  {Schwartz}},\ }\href@noop {} {\emph {\bibinfo {title} {{Quantum Field Theory
  and the Standard Model}}}}\ (\bibinfo  {publisher} {Cambridge University
  Press},\ \bibinfo {year} {2014})\BibitemShut {NoStop}%
\bibitem [{\citenamefont {Kittel}(2004)}]{Kittel2004}%
  \BibitemOpen
  \bibfield  {author} {\bibinfo {author} {\bibfnamefont {C.}~\bibnamefont
  {Kittel}},\ }\href
  {http://www.amazon.com/Introduction-Solid-Physics-Charles-Kittel/dp/047141526X/ref=dp_ob_title_bk}
  {\emph {\bibinfo {title} {Introduction to Solid State Physics}}},\ \bibinfo
  {edition} {8th}\ ed.\ (\bibinfo  {publisher} {Wiley},\ \bibinfo {year}
  {2004})\BibitemShut {NoStop}%
\end{thebibliography}%
%----------------------------------------------------------------------------------------
\end{document}